\documentclass[pra,aps,showkeys,nopacs,twocolumn,twoside]{revtex4}

\usepackage{amssymb}
\usepackage{amsmath}
\usepackage{amscd}
\usepackage{latexsym}  
\usepackage{epsfig}
\usepackage{bbm}
\usepackage{graphicx,epic,eepic}

\newtheorem{defi}{Definition}
\newtheorem{lemma}[defi]{Lemma}
\newtheorem{thm}[defi]{Theorem}
\newtheorem{cor}[defi]{Corollary}
\newtheorem{rem}[defi]{Remark}
\newtheorem{prop}[defi]{Proposition}
\newtheorem{exempel}[defi]{Example}

\newtheorem{conj}[defi]{Conjecture}

\def\e{\epsilon}

\newcommand{\qed}{\hfill $\Box$}
\newcommand{\tr}{{\operatorname{Tr}}}

\newcommand{\bra}[1]{{\langle{#1}|}}
\newcommand{\ket}[1]{{|{#1}\rangle}}
\newcommand{\ketbra}[1]{{\ket{#1}\!\bra{#1}}}
\newcommand{\C}{{\mathbbm{C}}}
\newcommand{\R}{{\mathbbm{R}}}

\newcommand{\E}{{\mathbbm{E}}}

\newcommand{\1}{{\openone}}

\newcommand{\csupp}{{\operatorname{csupp}\,}}
\newcommand{\esupp}{{\operatorname{esupp}\,}}
\newcommand{\qsupp}{{\operatorname{qsupp}\,}}

\newlength{\blank}
\newlength{\equalsign}
\newenvironment{beweis}[1][{\hspace{-\blank}}]{{\noindent\emph{Proof~{#1}.\ }}}{\hfill $\Box$\vskip 0.5\baselineskip}

\begin{document}

\title{Remote preparation of quantum states}
\date{1st June 2004}
\author{Charles H. Bennett}
\email{bennetc@watson.ibm.com}
\affiliation{IBM T.~J.~Watson Research Center, PO Box 218, Yorktown Heights, NY 10598, USA}
\author{Patrick Hayden}
\email{patrick@cs.caltech.edu}
\affiliation{Institute for Quantum Information, Caltech 107--81, Pasadena, CA 91125, USA}
\author{Debbie W. Leung}
\email{wcleung@cs.caltech.edu}
\affiliation{Institute for Quantum Information, Caltech 107--81, Pasadena, CA 91125, USA}
\author{Peter W. Shor}
\email{shor@research.att.com}
\affiliation{AT\&T Labs, Florham Park, NJ 07922, USA}
\author{Andreas Winter}
\email{winter@cs.bris.ac.uk}
\affiliation{Department of Computer Science, University of Bristol, Merchant Venturers Building,\\ Woodland Road, Bristol BS8 1UB, United Kingdom}

\begin{abstract}
  Remote state preparation is the variant of quantum state teleportation
  in which the sender knows the quantum state to be communicated.
  The original paper introducing teleportation established minimal
  requirements for classical communication
  and entanglement but the corresponding limits for remote state
  preparation have remained unknown until now: previous work has
  shown, however, that it not only requires less classical communication
  but also gives rise to a trade--off
  between these two resources in the appropriate setting.
  We discuss this problem from first principles,
  including the various choices one may follow in the definitions of the
  actual resources.
  \par
  Our main result is a general method of remote state preparation for arbitrary
  states of many qubits, at a cost of $1$ bit of classical communication and $1$ bit
  of entanglement per qubit sent.
  In this ``universal'' formulation, these ebit and cbit requirements
  are shown to be simultaneously optimal by exhibiting a dichotomy.
  Our protocol then yields the exact trade--off curve for memoryless
  sources of pure states
  (including the case of incomplete knowledge of the ensemble probabilities),
  based on the recently established quantum--classical trade--off
  for visible quantum data compression.
  A variation of that method allows us to solve the even more general problem
  of preparing entangled states between sender and receiver (i.e., purifications of
  mixed state ensembles).
  \par
  The paper includes an extensive discussion of our results, including
  the impact of the choice of model on the resources, the topic of obliviousness,
  and an application to private quantum channels and quantum data hiding.
\end{abstract}


\maketitle

\section{Introduction}
\label{sec:intro}
Teleportation~\cite{Teleportation} implements the
transmission of a quantum bit ($1$ qubit) by sending two classical bits
($2$ cbits), while using up quantum correlation amounting to one bit of
entanglement ($1$ ebit) -- although a description of this state would
require an infinite number of cbits, even when assisted by unlimited
classical correlation. What is more, in teleportation this description
is not needed at all: both the Sender and the Receiver act physically
on the state (i.e.~by quantum operations:
completely positive and trace preserving linear maps),
and the process can be used to transmit parts of entangled states
faithfully. This and the phenomenon of dense coding~\cite{Dense:coding} prove
that one cannot do with less than these resources: both $2$ cbits and $1$ ebit
are necessary.
\par
However, allowing the Sender knowledge of the state to be communicated changes
the task to what is now known as \emph{remote state preparation}
(r.s.p.)~\cite{Lo1999,Pati,Zeng:Zhang}, and here two new phenomena occur:
in~\cite{BDSSTW} it is shown that at the cost of possibly spending more entanglement
one can reduce the classical communication to $1$ cbit per qubit in the
asymptotics;
and there is a trade--off between the classical and the quantum resources needed,
of which \cite{BDSSTW} and~\cite{Devetak:Berger} provide bounds.
In the present work we put these results into their definite form by
proving a formula for the exact trade--off curve and by improving on the
result of~\cite{BDSSTW} to use only $1$ cbit and $1$ ebit per qubit.
\par
By a protocol for \emph{remote state preparation (r.s.p.)} we shall mean
a procedure involving two parties, a Sender who is given a description of
a state $\rho\in{\cal X}\subset{\cal S}({\cal K})$ from a subset ${\cal X}$
of the state set ${\cal S}({\cal K})$ of the Hilbert space ${\cal K}$,
and a Receiver who have access to a number of resources
(both forward and backward classical communication,
entanglement, shared randomness or others).
The protocol prescribes how to use these in a sequence of steps
(based on the previous exchange of messages in the protocol, and on $\rho$
for the Sender), resulting in a state $\widetilde{\rho}$ held by the
Receiver.
The dimension $D=\dim{\cal K}$ will, in the entire following
discussion be the principal asymptotic parameter (i.e., one should think
of it as large).
\par
We shall say that the protocol is \emph{(deterministic) exact} if
$\widetilde{\rho}=\rho$ for all choices of $\rho\in{\cal X}$.
\par
It is said to have \emph{fidelity $F$} if for all $\rho\in{\cal X}$,
$F(\widetilde{\rho},\rho)\geq F$, with the mixed--state
fidelity~\cite{Uhlmann:fidelity,Jozsa:fidelity}
$F(\rho,\sigma)=\|\sqrt{\rho}\sqrt{\sigma}\|_1^2
               =\tr\left(\sqrt{\sqrt{\rho}\sigma\sqrt{\rho}}\right)^2$.
(Note that for $F=1$ this is the same as an exact protocol.)
\par
A notion in between these two is a \emph{probabilistic exact} protocol
\emph{with error $\epsilon$}: this means that the protocol additionally
produces a flag, accessible to both Sender and Receiver, which indicates
``success'' or ``failure'' such that for all $\rho\in{\cal X}$,
$\Pr\{\text{``failure''}\}\leq\epsilon$ and $\widetilde{\rho}=\rho$
if the flag is ``success''; $\widetilde{\rho}$ is arbitrary otherwise.
(Note that such a protocol automatically has fidelity $\geq 1-\epsilon$.)
\par
Sometimes we want to impose a probability distribution $P$ on ${\cal X}$
and we will also consider protocols which have \emph{average fidelity
$\overline{F}$}, meaning
$$\int {\rm d}P(\rho)\, F(\widetilde{\rho},\rho) \geq \overline{F}.$$
\par
Varied as the parameters by which we judge the quality of a protocol are,
so are the ways to account for the use of resources: we will come back to
this issue later (subsection~\ref{subsec:resources}),
though the following example features not only
various quality measures, but also some choices of resource accounting.
For the moment we think only about protocols
which terminate at a certain prescribed point and the resources are those
needed to get to this point in the worst case $\rho\in{\cal X}$.
\begin{exempel}[Column method~\cite{BDSSTW}]
  \label{exp:column-method}
  The Sender is given an arbitrary pure state $\psi=\ketbra{\psi}$
  (note that we use \emph{state} synonymous with \emph{density operator};
  if we want to denote a state vector it will be $\ket{\psi}$)
  on a $D$--dimensional space
  (in~\cite{BDSSTW} $D=2^n$, i.e.~$n$ qubits), and that Sender and Receiver share
  sufficiently many maximally entangled states
  $\ket{\Phi_D}=\frac{1}{\sqrt{D}}\sum_{j=1}^D \ket{j}\ket{j}$
  of Schmidt rank $D$, labelled $1,2,\ldots,K$.
  \par
  The Sender performs the measurement
  $$\big( A_0=\overline{\psi}, A_1=\1-\overline{\psi} \bigr)$$
  on each of the $K$ entangled states and records the outcome.
  Here $\overline{\cdot}$ denotes the complex conjugation
  with respect to the basis $\{\ket{j}\}$ used to define $\Phi_D$.
  The probability
  of a $0$ clearly is $\frac{1}{D}$, the probability of a $1$ is $1-\frac{1}{D}$,
  hence the probability of $K$ $1$'s in a row (this will be called ``failure'') is
  $$\Pr\{\text{``failure''}\} = \left(1-\frac{1}{D}\right)^K
                                       \leq \exp\left(-\frac{K}{D}\right).$$
  \par
  Thus, if $\log K\geq \log D+\log\log\frac{1}{\epsilon}$,
  ``failure'' occurs with probability
  at most $\epsilon$. If this does not happen, there is at least one $0$ in the
  measurement results, and it requires $\log K$ cbits to communicate the label
  of the entangled state where it occurred to the Receiver. For definiteness, let
  us say that the Sender selects one position of outcome $0$ at random.
  Simple algebra shows that in this case the Receiver's reduced state is just $\psi$.
  \par
  This is an example of a probabilistic exact protocol with asymptotic cost of classical
  communication of $1$ cbit per qubit and success probability $1-\epsilon$.
  By ignoring the possibility of failure, it becomes a fidelity $1-\epsilon$ protocol.
  The protocol requires $K\log D$ ebits, which is exponential in the number of qubits.
  Most of this however can be recovered (``recycled'') using back communication after
  completion of the remote state preparation (see~\cite{BDSSTW}) such that only
  $O(\log D)$ ebits are irrecoverably lost.
  \par
  Clearly, to make this method deterministic exact, one must not put a limit on the
  number of trials $K$ (in which case the communication cost becomes infinite), or
  we must allow for a deterministic exact procedure in the case of ``failure'', e.g.
  teleportation.
  As this will increase the worst case communication cost to $2$ cbits per qubit,
  we are motivated to also consider \emph{expected} cbit cost, which in this example
  is $1+2\epsilon$ per qubit.
\end{exempel}
\par
As an aside to this exposition,
one can also consider making the task easier
for the Receiver, by only requiring that he is able to simulate
any measurement of which he is given a description, performed
on the state of which the Sender is given a description:
this is known as
\emph{classical teleportation}~\cite{CGM}, and though it is
related to our subject it lies outside the scope of the
present paper.
\par
The organisation of the rest of the paper is as follows:
in section~\ref{sec:universal} we present a general method of remote state
preparation, which uses $1$ cbit and $1$ ebit per qubit
asymptotically. It is based on an efficient state randomisation
method (see also~\cite{PQC}).
In section~\ref{sec:optimality} it is shown that any universal high--fidelity
protocol has to use $1$ cbit and $1$ ebit per qubit, asymptotically.
The cbit bound is true even if unlimited quantum back communication is allowed,
and the ebit bound is proved even in the presence of shared randomness.
We proceed to derive
the exact trade--off curve between ebits and cbits for an arbitrary ensemble
of candidate states, in section~\ref{sec:tradeoff}, using the recently
established analogous but simpler trade--off in quantum data compression
between qubits and cbits~\cite{HJW}.
Section~\ref{sec:ent} discusses the corresponding result
if ensembles of pure entangled states between the Sender and the Receiver
are to be prepared:
again, we can prove the exact trade--off between ebits and cbits.
\par
We conclude with a discussion of our findings and open questions in
section~\ref{sec:conclusion}: in particular considerations of the
issue of \emph{obliviousness} (cf.~\cite{Leung:Shor}) and a
discussion of the impact of certain slight changes in the
model on our conclusions.
\par
Several appendices contain separate or more technical issues:
in appendix~\ref{sec:gaussian} facts about Gaussian distributed vectors
are related; appendix~\ref{sec:randomisation} contains the proofs for
the central technical result, the state randomisation;
in appendix~\ref{sec:universal:qc-tradeoff} it is shown that universal description of
quantum states by qubits and cbits exhibits only a trivial trade--off between
the resources: there is a dichotomy between full quantum with no classical information
and no quantum with infinite classical information.
Facts about typical subspaces, used in various proofs,
are collected in appendix~\ref{app:types}.
Appendix~\ref{app:operational} contains thoughts on further operational
links between the qubit/cbit and the ebit/cbit trade--off, based on a conjecture
on the compressibility of mixed--state sources. Finally,
in appendix~\ref{app:proofs}, miscellaneous proofs are collected.
\par
Global notation conventions are: we use ${}^*$ for the Hermitian adjoint,
${}^\top$ for the transpose (in some given basis); $\exp$ and $\log$
are to basis $2$ (for the natural basis we use $e$, and the natural logarithm is
denoted $\ln$).


\section{Universal r.s.p.:\protect\\ 1 cbit + 1 ebit $\mathbf{\succ}$ 1 qubit}
\label{sec:universal}
We begin with a result on universal (approximate) state randomisation by unitaries:
\begin{thm}
  \label{thm:univ:randomisation}
  For Hilbert space ${\cal H}$ of dimension $D$ and $\epsilon>0$ there exist
  $$K \leq \left(\frac{10}{\epsilon}\right)^2
              \!D\,\log\left(\frac{20D}{\epsilon}\right)$$
  unitaries $U_k$ on ${\cal H}$ such that for every state $\varphi$,
  \begin{equation}
    \label{eq:univ:randomisation}
    \frac{1}{K}\sum_{k=1}^K U_k \varphi U_k^*
               \in \left[ \frac{1-\epsilon}{D}\1;\frac{1+\epsilon}{D}\1 \right],
  \end{equation}
  where the closed interval to the right refers to the operator order.
\end{thm}
\begin{beweis}
  Select the unitaries independently at random from the
  Haar measure on the unitary group. Observe that eq.~(\ref{eq:univ:randomisation})
  says that for all pure states $\varphi$ and $\psi$,
  \begin{equation*}
    \left| \frac{1}{K}\sum_{k=1}^K \tr\bigl( U_k \varphi U_k^* \psi \bigr)
                                                             - \frac{1}{D} \right|
    \leq \frac{\epsilon}{D}.
  \end{equation*}
  Fix a $\frac{\epsilon}{4D}$--net ${\cal M}$, according to lemma~\ref{lemma:net}.
  Lemma~\ref{lemma:conc} below allows us to bound
  \begin{equation*}\begin{split}
    \Pr&\left\{ \exists\varphi,\psi\!\in{\cal M}\ 
                  \left| \frac{1}{K}\sum_{k=1}^K \tr\bigl( U_k \varphi U_k^* \psi \bigr)
                                                                          - \frac{1}{D} \right|
                > \frac{\epsilon}{2D} \right\} \\
       &\phantom{=============:}
        \leq 2\left(\frac{20D}{\epsilon}\right)^{4D}\!\!
               \exp\left( -K\frac{\epsilon^2}{24} \right) \!.
  \end{split}\end{equation*}
  With triangle inequality for the trace norm we finally get
  \begin{equation*}\begin{split}
    \Pr&\left\{ \exists\varphi,\psi\ 
                 \left| \frac{1}{K}\sum_{k=1}^K \tr\bigl( U_k \varphi U_k^* \psi \bigr)
                                                                          - \frac{1}{D} \right|
               > \frac{\epsilon}{D} \right\} \\
       &\phantom{=============}
         \leq 2\left(\frac{20D}{\epsilon}\right)^{4D}\!\!
                \exp\left( -K\frac{\epsilon^2}{24} \right) \!,
  \end{split}\end{equation*}
  so if $K$ is as large as stated in the theorem there exist
  $U_1,\ldots,U_K$ such that eq.~(\ref{eq:univ:randomisation}) is true.
\end{beweis}
The probabilistic and geometrical facts used in the above proof are
contained in the following lemmas. The first is applied in the above proof
with $p=1$ but the general version is used later on.
\par
\begin{lemma}
  \label{lemma:conc}
  Let $\varphi$ be a pure state, $P$ a rank $p$ projector and let
  $(U_k)_{k=1}^K$ be an i.i.d. sequence of $U(D)$--valued
  random variables, distributed according to Haar measure. Then,
  for $0 < \epsilon \leq 1$,
  \begin{equation*}\begin{split}
    \Pr&\left\{ \left| \frac{1}{K}\sum_{k=1}^K \tr(U_k \varphi U_k^* P)
                                                         - \frac{p}{D} \right| 
        \geq \frac{\epsilon p}{D} \right\} \\
       &\phantom{===============} 
        \leq 2\exp\left( -Kp\frac{\epsilon^2}{6} \right).
  \end{split}\end{equation*}
\end{lemma}
\begin{beweis}
  In appendix~\ref{sec:randomisation}.
\end{beweis}
\begin{lemma}
  \label{lemma:net}
  Let ${\cal H}$ be a Hilbert space of dimension $D$. Then there exists,
  for every $\epsilon>0$, a set ${\cal M}$ of pure state vectors in ${\cal H}$
  of cardinality
  $$|{\cal M}| \leq \left(\frac{5}{\epsilon}\right)^{2D}$$
  such that for every state vector $\ket{\varphi}\in{\cal H}$ there exists
  a state vector $\ket{\psi}\in{\cal M}$ such that
  \begin{equation*}
    \bigl\| \varphi-\psi \bigr\|_1
          \leq 2\sqrt{1-F(\varphi,\psi)}
          \leq 2\bigl\| \ket{\varphi}-\ket{\psi}\bigr\|_2
          \leq \epsilon.
  \end{equation*}
  Such a set ${\cal M}$ we call \emph{$\epsilon$--net}.
\end{lemma}
\begin{beweis}
  In appendix~\ref{sec:randomisation}.
\end{beweis}
A few words of interpretation: it is known~\cite{AMTdW,boykin:roychowdhury}
that if $\epsilon=0$, one needs $K\geq D^2$, and this is tight as the example
of the generalised Pauli (sometimes called Weyl) operators shows.
We call a selection of unitaries as in the theorem
``randomising'', because application of a randomly chosen $U_k$ results in an
almost maximally mixed state. Clearly, this has cryptographic applications,
an exploration of which is to be found in our separate paper~\cite{PQC}.
\par
Let us now show how to use this result to build a remote state preparation
protocol: first of all, given a pure state $\psi$, one can write down the
family of operators
\begin{align*}
  A_k             &= \frac{D}{K(1+\epsilon)} U_k \overline{\psi} U_k^*
                                                             \quad (k=1,\ldots,K) \\
  A_{\rm failure} &= \1-\sum_{k=1}^K A_k.
\end{align*}
This is a POVM by virtue of theorem~\ref{thm:univ:randomisation}.
\par\medskip\noindent
{\bf Protocol $\mathbf{\Pi}$ (Description of $\psi$ at the Sender):}
\begin{enumerate}
  \item The Sender measures the POVM $(A_k)$ of the above description
    on her half of the entangled state $\Phi_D$.
    and announces the result (either ``failure'' or $k=1,\ldots,K$).
  \item If the message received is not ``failure'', say $k$, the Receiver
    applies the unitary $U_k^\top$ to his part of the state
    $\Phi_D$.
\end{enumerate}
\begin{thm}
  \label{thm:univ:rsp}
  The above protocol realises remote state preparation for an
  arbitrary state $\ket{\psi}\in{\cal K}$ exactly with a probability
  of failure of exactly $\frac{\epsilon}{1+\epsilon} \leq \epsilon$.
  \par
  In particular, exact probabilistic r.s.p.~with error $\epsilon$
  is possible using
  \begin{align*}
    \log D+2\log\frac{10}{\epsilon}+\log\log\frac{20D}{\epsilon} & \ \text{ cbits}  \\
    \text{and}\phantom{===========:}\log D                       & \ \text{ ebits.}
  \end{align*}
\end{thm}
\begin{beweis}
  It is straightforward to check that the protocol, in case it does not
  produce a failure, exactly prepares $\ket{\psi}$ at the Receiver.
  \par
  For the probability assertions: the event $k$ of the POVM $(A_k)$ is
  triggered with probability exactly $\frac{1}{K(1+\epsilon)}$. Hence the
  probability of failure is
  $$1-K\frac{1}{K(1+\epsilon)} = 1-\frac{1}{1+\epsilon} = \frac{\epsilon}{1+\epsilon}.$$
  \par
  The remaining claims are easy consequences of this.
\end{beweis}
\begin{cor}
  \label{cor:1c1e}
  Probabilistic exact remote state preparation is possible with
  $1$ cbit and $1$ ebit per qubit, asymptotically.
  \qed
\end{cor}

\section{Optimality of cbit and\protect\\ ebit resources}
\label{sec:optimality}
We will now show that both $1$ ebit and $1$ cbit per qubit are necessary asymptotically
for universal r.s.p.~protocols with high fidelity. More precisely, we assume a protocol
like our protocol $\Pi$ in section~\ref{sec:universal}, which takes as input the description
of an arbitrary state $\psi$ on a $D$--dimensional space ${\cal K}$, uses an entangled state
of Schmidt rank $S$, forward communication of one out of $K$ messages, such that
the output states $\widetilde{\rho}$ have fidelity $F$ to the ideal $\psi$.
\par
Regarding the communication resources, causality shows that $K\geq FD$ is necessary,
even if unlimited \emph{quantum} back communication is allowed: this is
because the mere capability to remotely prepare an orthogonal basis of states with fidelity
$F$ clearly allows the Sender to transmit one out of $D$ classical messages with
probability at least $F$ of correct decoding.
Imagine now that Sender and Receiver follow the r.s.p.~protocol with the modification
that each forward communication is skipped and replaced by the Receiver guessing it
at random.
\par
In this modification of the protocol, the probability of correct decoding clearly
is $\geq \frac{F}{K}$, as the Receiver has only to guess the correct classical
communication out of $K$.
But the modified protocol involves no forward transmission at all, hence
the probability of correctly identifying the Sender's message --- $1$ out of $D$ ---
is $\leq \frac{1}{D}$: this shows $\frac{1}{D} \geq \frac{F}{K}$.
\par
We have thus proved:
\begin{thm}
  \label{thm:causality}
  Any r.s.p.~protocol with fidelity $F$ requires classical communication of
  $$C=\log K\geq \log D+\log F\text{ cbits,}$$
  even if unlimited quantum back communication is allowed.
  \qed
\end{thm}
\par
Regarding the entanglement, we have the following result of an extremely strong dichotomy:
\begin{thm}
  \label{thm:dichotomy}
  Any r.s.p.~protocol using an entangled state of Schmidt rank
  $S\leq qD$ ($q<F$) requires classical communication of
  $$C\geq \frac{q(1-q)}{6}D-O(\log D)\text{ cbits,}$$
  even if unlimited shared randomness is available.
  \par
  On the other hand, there is a protocol with fidelity
  $F\geq 1-\epsilon$, which uses \emph{no entanglement at all}
  (i.e., $S=1$), and classical communication of
  $$C\leq \left( 4+\log\frac{1}{\epsilon} \right) D \text{ cbits.}$$
\end{thm}
Thus, in the asymptotic limit, and with
normalised resources $E=\log S/\log D$ and $R=C/\log D$
for the entanglement and communication rates,
the rate point $E=1$ marks the threshold between
two drastically different regimes: for $E\geq 1$, the classical communication rate
$R=1$ is sufficient by corollary~\ref{cor:1c1e} and necessary by
theorem~\ref{thm:causality}. For any entanglement rate $E<1$,
theorem~\ref{thm:dichotomy} shows that no finite classical communication rate
is possible: $R\rightarrow\infty$ with $D\rightarrow\infty$.
\par
Thus, $E\geq 1$ and $R\geq 1$ hold simultaneously
and both equalities can be achieved at the same time (theorem~\ref{thm:univ:rsp}),
unless $R=\infty$ in which case $E=0$.
I.e., there is only a trivial trade--off between ebits and cbits.
\par\medskip\noindent
\begin{beweis}[of theorem~\ref{thm:dichotomy}]
  Consider any protocol, using a shared random variable $\nu$, so that
  the output state $\widetilde\rho$ is the mixture of the output states
  for the various values of $\nu$.
  Such a protocol clearly has average fidelity $\overline{F}\geq F$, with
  respect to the uniform (i.e., unitarily invariant) distribution
  on the pure states:
  $$\overline{F} = \int {\rm d}\ket{\psi}\, F(\widetilde{\rho},\ketbra{\psi}).$$
  Because of the linearity of the pure state fidelity in $\widetilde\rho$,
  $\overline{F}$
  is the probabilistic average of the fidelities $\overline{F}_\nu$
  of the protocol for the value $\nu$ of the shared random variable.
  Hence there exists a $\nu$ such that $\overline{F}_\nu \geq F$,
  and we can consider a new protocol, without shared randomness,
  which has the same fidelity as the original.
  \par
  Thus, w.l.o.g., we may assume a protocol of the form
  described in the first paragraph
  of this section, which uses only the entangled state $\Phi$ and
  forward classical communication.
  In general terms, it proceeds by the Sender performing a measurement
  on her half of $\Phi$ and communicating the outcome $m$ to the Receiver,
  who then applies a quantum operation $T_m$ to his half of $\Phi$.
  Observe that after the Sender's measurement the state of the Receiver is collapsed
  to a state \emph{supported on the support of the restriction of $\Phi$},
  which is a space of dimension $S$. Thus, effectively, the Sender supplies
  the Receiver with a message $m$ and a state $\xi$ on an $S$--dimensional
  system, from the combination of which an approximation of $\psi$
  is obtained: $\widetilde{\rho}=T_m(\xi)$. Once more using bilinearity of the pure state
  fidelity, we may assume that the choice of the pair $(m,\xi)$ from
  $\psi$ is deterministic, and that $\xi$ is a pure state.
  (This no longer describes an r.s.p. protocol, where uncontrollable
  randomness due to measurements is the rule: what is important here is
  that this can only enhance the capabilities of the Sender.)
  \par
  We now invoke theorem~\ref{thm:univ:description} from
  appendix~\ref{sec:universal:qc-tradeoff},
  which lower bounds the classical communication cost
  of such a quantum--classical state description: we obtain
  $$C \geq \frac{q(1-q)}{6} D-O(\log D),$$
  which is our claim.
  \par
  Conversely, in the situation with no entanglement, pick an $\sqrt{4\epsilon}$--net
  ${\cal M}$ of cardinality at most
  $\left( \frac{5}{2\sqrt{\epsilon}} \right)^{2D}$,
  according to lemma~\ref{lemma:net}. Clearly, a valid protocol is this:
  \par
  Given a state description of $\psi$, the Sender picks a $\ket{\phi}\in{\cal M}$
  with fidelity $1-\epsilon$ (because lemma~\ref{lemma:net} is strong enough for that)
  to $\psi$, and sends the Receiver an identifier for $\phi$,
  which requires $\log|{\cal M}|$ cbits.
\end{beweis}

\section{Ensemble trade--off curve}
\label{sec:tradeoff}
While in the previous sections we considered universal r.s.p.
(even though asymptotic, allowing \emph{any} input state), in the present and
following section we want to look at \emph{ensemble asymptotics}:
we consider an ensemble of quantum states ${\cal E}=\{\ket{\psi_i},p_i\}$ on the Hilbert
space ${\cal H}$ of dimension $d$, and are interested in r.s.p. of the
ensemble $\{\ket{\psi_I},p_I\}$ on ${\cal H}^{\otimes n}$, with states
and probabilities
\begin{align*}
  \ket{\psi_I} &= \ket{\psi_{i_1}}\otimes\cdots\otimes\ket{\psi_{i_n}}, \\
  p_I          &= p_{i_1}\cdots p_{i_n},                        \\
  I            &= i_1\ldots i_n,
\end{align*}
and for large $n$. The notation for letters (lower case) and blocks (upper case)
is used throughout this and the following section.
\par
Note that even in the case that the ensemble contains
\emph{all} pure states on ${\cal H}$, the asymptotics will capture only
the product states in ${\cal K}={\cal H}^{\otimes n}$, unlike the
model of the previous sections.
\par
We shall be interested in protocols which have average fidelity
$\overline{F}$, i.e.,
\begin{equation}
  \label{eq:global-fidelity}
  \sum_I p_I \tr\bigl(\ketbra{\psi_I}\widetilde{\rho}_I\bigr) \geq \overline{F}.
\end{equation}
By the monotonicity of the fidelity under partial traces, this implies the
weaker condition
\begin{equation}
  \label{eq:local-fidelity}
  \sum_I p_I \frac{1}{n}\sum_{k=1}^n
                 \tr\bigl(\ketbra{\psi_{i_k}}\tr_{\neq k}\widetilde{\rho}_I\bigr)
                                                                    \geq \overline{F},
\end{equation}
which we will find useful at times.
\par
Note that by considering average fidelities as we do here, shared
randomness becomes automatically useless, because we aim to prepare pure states
with high fidelity (compare the proof of theorem~\ref{thm:dichotomy}).
\par
On block--length $n$, a protocol for r.s.p.~uses a maximally
entangled state $\Phi_D$ of Schmidt rank $D$ shared between Sender (A)
and Receiver (B).
We consider here protocols which use only forward communication:
their general form is described by a measurement POVM depending on $I$,
${\bf M}^I=(M^I_j)_{j}$ with $j$ running over a set $\{1,\ldots,K\}$: after performing
this POVM on her half of $\Phi_D$, the Sender communicates $j$, and
the Receiver applies a quantum operation $T_j$ to his half of $\Phi_D$.
We write $({\bf M},T)$ to denote such a protocol, sometimes adding a subscript
$n$ to indicate the block--length.
\par
The resources used are defined, in a way similar to~\cite{HJW}, as
the entanglement rate
$$\esupp({\bf M},T) := \frac{1}{n}\log D,$$
and the communication rate
$$\csupp({\bf M},T) := \frac{1}{n}\log K.$$
(The notation is meant to remind one of ``support'', since what
we count here is the number of bits necessary to support the
entanglement and the classical messages, respectively.)
We say that a rate pair $(R,E)$ is \emph{achievable} if for all $\epsilon,\delta>0$
there exists $n_0$ such that for all $n\geq n_0$ there are r.s.p. protocols
$({\bf M},T)_n$ with fidelity $1-\epsilon$ and resources
\begin{align*}
  \csupp({\bf M},T) &\leq R+\delta, \\
  \esupp({\bf M},T) &\leq E+\delta.
\end{align*}
This allows us to rigorously define the trade--off function $E^*$ by
$$E^*(R)=\min\{ E | (R,E) \text{ is achievable} \}.$$
\par
A similar trade--off is studied in~\cite{HJW} between cbits and transmitted
qubits instead of ebits, which is a visible coding generalisation of
the familiar Schumacher quantum data compression~\cite{Schumacher,Schumacher:Jozsa}:
such a protocol consists of a pair $(E_n,D_n)$ of encoding and decoding maps.
The encoding takes $I$ to a combination of a quantum message supported on
$n\,\qsupp(E_n,D_n)$ qubits and a classical message comprising $n\,\csupp(E_n,D_n)$
cbits, while the decoding is a quantum operation acting on these two,
with the aim as before, to achieve a large average input--output fidelity.
\par
Defining achievable rate pairs $(R,Q)$ analogous to the above, and letting
$$Q^*(R)=\min\{ Q | (R,Q) \text{ is achievable} \},$$
we have the following single--letter formula for the
quantum--classical trade--off (q.c.t.) curve:
\begin{thm}[Hayden,~Jozsa~and~Winter~\cite{HJW}]
  \label{thm:qct}
  \begin{equation}
    \label{eq:qct}
    Q^*(R)=M({\cal E},R):=\min\left\{ S(A:B|C) | S(A:C)\leq R \right\},
  \end{equation}
  where the minimisation is over all tripartite states
  \begin{equation}
    \label{eq:omega}
    \omega=\sum_i p_i\ketbra{i}^A\otimes\psi_i^B\otimes\sum_j p(j|i)\ketbra{j}^C,
  \end{equation}
  for stochastic matrices $p(j|i)$; $j$ has a range of at most $m+1$
  if the ensemble consists of $m$ states.
  \begin{align*}
    S(A:C)   &= S(A)+S(C)-S(AC)\text{ and}\\
    S(A:B|C) &= S(AC)+S(BC)-S(ABC)-S(C),
  \end{align*}
  are the (conditional) quantum mutual information, defined
  via the von Neumann entropy $S$, referring
  implicitely to the state $\omega$: $S(AC)$ is the von Neumann
  entropy of $\omega$ restricted to $AC$, etc.
  \qed
\end{thm}
In brief, once an optimal channel $p(j|i)$ is chosen, the scheme essentially
works as sending part of the classical encoding $J=j_1\ldots j_n$ (only typical)
using the Reverse Shannon Theorem~\cite{BSST}, and then
Schumacher--compressing the induced ``conditional'' ensemble
\begin{align*}
  \{\psi_I,q(I|J) &=q(i_1|j_1)\cdots q(i_n|j_n)\}, \text{ with}      \\
  q(i|j)          &=\left( \sum_i p_i p(j|i) \right)^{-1} p_i p(j|i),
\end{align*}
to its von Neumann entropy (note that the ensemble is a product of
independent ensembles even though they are not all identical).
\par
For each point $(R,Q)$ on the trade--off curve for the ensemble ${\cal E}$
we can, with the method of the previous section, construct an asymptotic
and approximate r.s.p.~protocol using $C=R+Q$ cbits and $E=Q$ ebits:
We only have to use theorem~\ref{thm:univ:rsp} to remotely
prepare the encoded state on $Q$ qubits, using $Q$ ebits
and an additional $Q$ cbits, all per qubit.
\par
We can summarise the finding as an upper bound on $E^*(R)$, in a strange
implicit form:
\begin{equation}
  \label{eq:E:upperbound}
  E^*(R+Q^*(R))\leq Q^*(R).
\end{equation}
\begin{rem}
  Devetak and Berger~\cite{Devetak:Berger} happened to parametrise the q.c.t.~curve
  for the uniform qubit ensemble. Using teleportation instead of our
  theorem~\ref{thm:univ:rsp} they obtained r.s.p.~protocols using
  $C+2Q$ cbits and $Q$ ebits.
\end{rem}
Using the chain rule $S(A:B|C)+S(A:C)=S(A:BC)$,
we can put together theorem~\ref{thm:qct}
and eq.~(\ref{eq:E:upperbound}) to obtain that
$$E^*(R) \leq \min\{ S(A:B|C) | S(A:BC)\leq R \}.$$
In fact, we shall show in a moment that equality holds here:
\begin{thm}
  \label{thm:rsp}
  \begin{equation}
    \label{eq:rsp}
    E^*(R)=N({\cal E},R):=\min\{ S(A:B|C) | S(A:BC)\leq R \},
  \end{equation}
  where the minimisation is over all tripartite states $\omega$
  as in eq.~(\ref{eq:omega}).
\end{thm}
Before we prove this, we state a little lemma collecting some properties of $N$:
\begin{lemma}
  \label{lemma:N:properties}
  $N$ is convex, continuous and strictly decreasing in the interval
  where it takes finite positive values, which is $[S(B);S(A)]$.
  It obeys the following additivity relation for
  ensembles ${\cal E}_1$ and ${\cal E}_2$:
  \begin{equation}
    \label{eq:N:add}
    N({\cal E}_1\otimes{\cal E}_2,R)
         = \min\bigl\{ N({\cal E}_1 \!,\! R_1)
                      +N({\cal E}_2 \!,\! R_2) | R_1+R_2 \!=\! R \bigr\} \!.
  \end{equation}
\end{lemma}
\begin{beweis}
  In appendix~\ref{app:proofs}.
\end{beweis}
\begin{beweis}[of theorem~\ref{thm:rsp}]
  Only the direction ``$\geq$'' has to be proved:
  assume an r.s.p.~protocol for block--length $n$ and with
  average fidelity
  $$\overline{F}=    \sum_I p_I\tr\bigl(\ketbra{\psi_I}\widetilde{\rho_I}\bigr)
                \geq 1-\epsilon.$$
  Let us describe the protocol again: the Sender performs a measurement
  on her half of $n(E+\delta)$ EPR--pairs and sends the measurement result $j$
  (obtained with probability $p(J|I)$ and collapsing the Receiver's
  state to $\sigma_{I,j}$)
  to the Receiver (using $n(R+\delta)$ classical bits), who performs a quantum
  operation $T_j$ on his half of the EPR--pairs. The state thus produced
  is $\widetilde{\rho}_{I,j}$ and obviously
  $\widetilde{\rho}_I=\sum_j p(j|I)\widetilde{\rho}_{I,j}$.
  \par
  Now, the post--measurement state, including a classical system $A$
  to record $I$, can be written in the general form
  $$\sigma=\sum_I p_I\ketbra{I}^A \otimes
               \sum_j p(j|I) \sigma_{I,j}^{B}\otimes\ketbra{j}^C,$$
  where $C$ is the classical system used for communicating $j$.
  \par
  Entropic quantities of this state are related to the resources
  required by the protocol:
  first of all, $S_\sigma(A:BC)\leq n(R+\delta)$ because in total $n(R+\delta)$ bits
  are communicated, and their information cannot be exceeded by the
  information in what the receiver eventually gets, by causality.
  Similarly, because all the $\sigma_{I,j}^B$ are supported on the
  $nE$ qubits which form the Receiver's half of the EPR--pairs,
  we get
  $$n(E+\delta) \geq S_\sigma(B) \geq S_\sigma(B|C) \geq S_\sigma(A:B|C).$$
  \par
  We may assume that the $T_j$ do not affect the system $C$, and because
  (conditional) mutual informations are non--increasing under local
  quantum operations, we obtain that
  \begin{align}
    \label{eq:R-lower}
    n(R+\delta) &\geq S_{\widetilde{\rho}}(A:BC),  \\
    \label{eq:E-lower}
    n(E+\delta) &\geq S_{\widetilde{\rho}}(A:B|C),
  \end{align}
  with the state
  $$\widetilde{\rho}=\sum_I p_I\ketbra{I}^A \otimes
                   \sum_j p(j|I) \widetilde{\rho}_{I,j}^{B}\otimes\ketbra{j}^C.$$
  (Note that $\widetilde{\rho}_{I,j}=T_j(\sigma_{I,j})$ for all $I,j$.)
  Our goal is now to switch in the latter expression to the ideal
  states $\ketbra{\psi_I}$, arguing that we retain high fidelity to
  $\widetilde{\rho}$, and then invoking general continuity bounds
  for the entropy:
  \par
  More precisely, define
  $$\Omega=\sum_I p_I\ketbra{I}^A \otimes
                   \sum_j p(j|I) \ketbra{\psi_I}^{B}\otimes\ketbra{j}^C.$$
  Then we can estimate
  \begin{equation*}\begin{split}
    \bigl\| \Omega-\widetilde{\rho} \bigr\|_1
          &=    \sum_I p_I \sum_j p(j|I)
                     \bigl\| \ketbra{\psi_I}-\widetilde{\rho}_{I,j} \bigr\|_1 \\
          &\leq \sum_I p_I \sum_j p(j|I)
                     2\sqrt{ 1-\tr(\ketbra{\psi_I}\widetilde{\rho}_{I,j}) }   \\
          &\leq \sum_I p_I 2\sqrt{ 1-\tr(\ketbra{\psi_I}\widetilde{\rho}_I) } \\
          &\leq 2\sqrt{1-\overline{F}}
           \leq 2\sqrt{\epsilon},
  \end{split}\end{equation*}
  where in the second line we have used the inequality
  $\frac{1}{2}\|\rho-\sigma\|_1\leq \sqrt{1-F(\rho,\sigma)}$ for states
  $\rho,\sigma$~\cite{Fuchs:vandeGraaf}, and then concavity of the square
  root function.
  Because for states $\rho,\sigma$ on a $D$--dimensional system,
  $\|\rho-\sigma\|_1\leq\epsilon\leq\frac{1}{2}$, we have the
  Fannes inequality~\cite{fannes}
  $|S(\rho)-S(\sigma)|\leq -\epsilon\log\frac{\epsilon}{D}$,
  we obtain that there exists a function $f(\epsilon)$,
  vanishing as $\epsilon\rightarrow 0$, such that
  \begin{align}
    \label{eq:almost-done:R}
    n(R+\delta) &\geq S_\Omega(A:BC)-n f(\epsilon), \\
    \label{eq:almost-done:E}
    n(E+\delta) &\geq S_\Omega(A:B|C)-n f(\epsilon).
  \end{align}
  The reasoning is that the entropies of combinations of $A$, $B$ and $C$
  relative to the states $\Omega$ and $\widetilde{\rho}$, see
  eqs.~(\ref{eq:R-lower}) and (\ref{eq:E-lower}), can be estimated
  against each other by the Fannes inequality, observing that Hilbert space
  dimensions are of the form $X^n$ with a constant $X$.
  \par
  Hence we get (letting $\delta'=\delta+f(\epsilon)$)
  \begin{equation*}\begin{split}
    n(E+\delta') &\geq \min_\Omega\{ S(A:B|C) | S(A:BC)\leq n(R+\delta') \} \\
                 &=    N\bigl( {\cal E}^{\otimes n},n(R+\delta') \bigr)
  \end{split}\end{equation*}
  Now we invoke lemma~\ref{lemma:N:properties} to estimate further,
  \begin{equation*}\begin{split}
    E+\delta' &\geq \frac{1}{n} N\bigl( {\cal E}^{\otimes n},n(R+\delta') \bigr) \\
              &=    \min\left\{ \frac{1}{n}\sum_{k=1}^n N({\cal E},R_k)
                                 \left| \frac{1}{n}\sum_{k=1}^n R_k = R+\delta' \right.\right\} \\
              &\geq N({\cal E},R+\delta'),
  \end{split}\end{equation*}
  the second line by eq.~(\ref{eq:N:add}), the third by convexity
  of $N$. Using the continuity of $N$ with $\delta'\rightarrow 0$
  (which occurs with $\epsilon,\delta\rightarrow 0$), we arrive at
  $E \geq N({\cal E},R)$, as desired.
\end{beweis}
\par
Readers of~\cite{HJW} will notice the similarity of the proofs
of the lower bounds in theorems~\ref{thm:qct} and~\ref{thm:rsp}.
Given that for the upper bound we use an operational transformation of
a q.c.t.~protocol into an r.s.p.~protocol,
one may wonder if there is not a proof of the optimality
of this reduction by an inverse reduction of an r.s.p.~protocol
to a q.c.t.~protocol. We relate one such attempt in
appendix~\ref{app:operational}.
\par
There are two generalisations of theorem~\ref{thm:qct} which we can transport
to obtain more general versions of theorem~\ref{thm:rsp}:
the first is to lift the restriction to discrete ensembles, which is not really
necessary - it is shown in~\cite{HJW} by suitable approximation (using in fact
the net lemma~\ref{lemma:net}) that theorem~\ref{thm:qct} holds true for
an arbitrary probability distribution $p$ on the pure states of ${\cal H}$.
This shows automatically that theorem~\ref{thm:rsp} also holds in the same form
for general ensembles (in general with $\inf$ instead of $\min$).
\par
The second concerns the so--called \emph{arbitrarily varying sources (AVS)}:
an ensemble is generally taken to represent some partial knowledge about the
states to be encountered, and this model allows us to fine--tune this to even less
knowledge: an AVS is a family of probability distributions $p_s$, $s\in{\cal S}$
on the space of pure states, with the intention that at each time step each of
the distributions $p_s$ can occur. One might want to think of an adversary
choosing $s^n=s_1\ldots s_n$, thus presenting a given protocol with the
distribution of states
$$p_{s^n}=p_{s_1}\otimes\cdots\otimes p_{s_n}.$$
A protocol (of either q.c.t.~or r.s.p.) is said to have fidelity $\overline{F}$
if for all choices $s^n\in{\cal S}^n$,
$$\int {\rm d}p_{s^n}(\psi) F(\ketbra{\psi},\widetilde{\rho}) \geq \overline{F},$$
where $\widetilde{\rho}$ is the output state on input $\psi$.
\par
It turns out~\cite{HJW} that for q.c.t.~there is still a trade--off
in this case, and that $Q^*(R)$ is given by the trade--off for the
worst--case ensemble distribution from the convex hull
${\bf P}={\rm conv}\{p_s|s\in{\cal S}\}$ of the $p_s$:
\begin{thm}
  \label{thm:qct-avs}
  For an AVS $\{p_s\}_{s\in{\cal S}}$, the q.c.t.~trade--off curve
  is given by
  $$Q^*(R)=\sup_{p\in{\bf P}} Q^*(p,R),$$
  where $Q^*(p,R)$ is the trade--off of theorem~\ref{thm:qct}
  as a function of cbit rate $R$ and the ensemble distribution $p$,
  made explicit.
  \qed
\end{thm}
This immediately implies, by the same reasoning, the corresponding
theorem for remote state preparation:
\begin{thm}
  \label{thm:rsp-avs}
  For an AVS $\{p_s\}_{s\in{\cal S}}$, the r.s.p.~trade--off curve
  is given by
  $$E^*(R)=\sup_{p\in{\bf P}} E^*(p,R),$$
  where $E^*(p,R)$ is the trade--off of theorem~\ref{thm:rsp}
  as a function of cbit rate $R$ and the ensemble distribution $p$,
  made explicit.
  \qed
\end{thm}
\par
\begin{figure}[ht]
  \includegraphics[width=8.5cm]{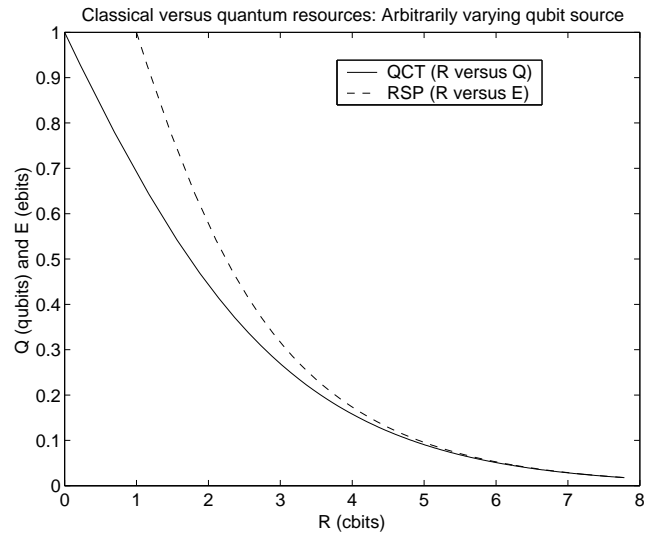}
  \caption{The q.c.t. trade--off curve for qubits vs.~cbits according to
           Devetak and Berger~\cite{Devetak:Berger} (solid)
           and the implied r.s.p. trade--off for ebits vs.~cbits (broken).}
  \label{fig:qubit:tradeoff}
\end{figure}

In particular, dropping all restrictions, i.e.~for the AVS
with ${\bf P}=\{\text{all distributions}\}$ (which means that
the adversary may pick an arbitrary product state for the protocol),
we obtain the ``ultimate''
trade--off functions ${\bf Q}^*$ and ${\bf E}^*$:
these govern the asymptotic qubit/cbit and ebit/cbit cost of
compressing and remotely preparing blocks of arbitrary states.
Because we know that $Q^*(R)$ for the uniform distribution
dominates all other curves with fixed input distribution
(\cite{HJW}, theorem 6.1 and corollary 9.2),
we have ${\bf Q}^*(R)=Q^*(R,\text{uniform})$ and hence
${\bf E}^*(R)=E^*(R,\text{uniform})$. For qubits we thus can plot
${\bf E}^*$ thanks to the results of Devetak and Berger~\cite{Devetak:Berger}
(Fig.~\ref{fig:qubit:tradeoff}).
\par
\par\medskip
A word might be necessary to explain why there is no contradiction between this
universal trade--off curve (which evidently exists not just
for qubits, but for any qu\emph{d}its; to our knowlegde, however, it
hasn't been worked out explicitly for $d>2$),
and the proof of the nonexistence of any finite trade--off in section~\ref{sec:optimality}.
This is because in
the present section the task is much less ambitious: we only want to remotely prepare
large blocks of (admittedly arbitrary) qubit states, i.e. a long product of pure states
in small dimension. The set of product states however is much smaller than the set
of all pure states on the large blocks. This fact is sufficient to allow
an efficient trade--off between ebits (or qubits) and cbits.

\section{Preparation of entangled states}
\label{sec:ent}
It is tempting to consider the generalisation of the previous section to
\emph{mixed state sources}. Observing however that our solution of the pure state
case rested on the quantum--classical trade--off for pure state
compression~\cite{HJW} --- itself a generalisation of Schumacher's
source coding~\cite{Schumacher} --- we might be discouraged by the corresponding
mixed--state compression being far from resolved.
A glimpse of this is
provided in appendix~\ref{app:operational}, but see a more detailed
discussion in~\cite{BCFJS,Jozsa:Winter} and references therein.
\par
Instead, we target a seemingly harder problem: the Sender (A) should remotely prepare
an entangled state between the Receiver (B) and herself, drawn from an ensemble.
Clearly, the Receiver in this way obtains the mixed state ensemble of the
reduced states.
\par
In detail, assume an ensemble
${\cal E}=\{ \ket{\varphi_i}^{AB},p_i \}_{i=1}^m$
of pure entangled states generating the i.i.d.~source
\begin{align*}
  I               &= i_1\ldots i_n,                                              \\
  \ket{\varphi_I} &= \ket{\varphi_{i_1}}\otimes\cdots\otimes\ket{\varphi_{i_n}}, \\
  p_I             &= p_{i_1}\cdots p_{i_n}.
\end{align*}
The protocols we consider are of a general form very similar to those
in section~\ref{sec:tradeoff}: they allow both parties to use a maximally
entangled state $\Phi_D$ of Schmidt rank $D$, and consist of a family of
instruments~\cite{davies:lewis} ${\bf M}^I=(M_j^I)_j$ ($j=1,\ldots,M$)
for the Sender, i.e.~each $M_j^I$ is a completely positive map,
and their sum (over $j$) is a trace preserving map for every $I$ --- this
conveniently captures the notion of a (partial) measurement with a post--measurement
state. Furthermore, there are quantum operations $T_j$ for the Receiver.
The states prepared in this way are
$$\widetilde{\rho}_I = \sum_j (M_j^I\otimes T_j)\Phi_D,$$
and as before we demand that the fidelity $\overline{F}\geq 1-\epsilon$, with
$$\sum_I p_I \tr\bigl( \varphi_I\widetilde{\rho}_I \bigr) \geq \overline{F}.$$
And similarly, we call a rate pair $(R,E)$ \emph{achievable} if for
all $\epsilon,\delta>0$ and sufficiently large $n$ there exist
r.s.p.~protocols with
\begin{align*}
  \frac{1}{n}\log M &\leq R+\delta, \\
  \frac{1}{n}\log D &\leq E+\delta.
\end{align*}
Define the trade--off function for the ensemble ${\cal E}$,
$$E^*(R) = \min\{ E | (R,E) \text{ achievable} \}.$$
\par
We start by describing a protocol to achieve the rate point
with the smallest $R$ allowed by causality (a different proof
for the achievability of the cbit rate can be found in~\cite{berry:sanders}
even though with a method that is very wasteful in terms of entanglement,
much like the column method of example~\ref{exp:column-method}):
\begin{prop}
  \label{prop:peters:protocol}
  There exists an r.s.p.~protocol which achieves the rate pair
  \begin{align*}
    R &= \chi\bigl( \{ p_i,\varphi_i^B \} \bigr), \\
    E &= S\left(\sum_i p_i\varphi_i^B\right),
  \end{align*}
  with the Holevo quantity $\chi$ of the Receiver's
  mixed state ensemble $\{p_i,\varphi_i^B\}$.
\end{prop}
\begin{beweis}
  Consider a string $I=i_1\ldots i_n$ of type (i.e. relative letter frequencies)
  $Q$ --- see appendix~\ref{app:types} for details ---, and construct
  (with $\delta>0$)
  the conditional typical projector $\Pi^n_{\varphi^B\!,\delta}(I)$
  for $\varphi^B_{I}=\varphi^B_{i_1}\otimes\cdots\otimes\varphi^B_{i_n}$.
  By eq.~(\ref{eq:c-typical:prob}), for sufficiently large $n$,
  $$\tr\bigl( \varphi^B_{I}\Pi^n_{\varphi^B\!,\delta}(I) \bigr) \geq 1-\epsilon.$$
  Construct also the typical projector $\Pi:=\Pi^n_{\rho,\delta}$
  of the average state $\rho=\sum_i Q(i)\varphi^B_i$:
  by lemma~\ref{lemma:law:large-numbers}, for sufficiently large $n$,
  $$\tr\bigl( \varphi^B_{I}\Pi^n_{\rho,\delta} \bigr) \geq 1-\epsilon.$$
  Hence, if we define (for all $I$ of type $Q$)
  $$\pi_{I} := \Pi^n_{\rho,\delta}\Pi^n_{\varphi^B\!,\delta}(I)
                      \varphi^B_{I}
               \Pi^n_{\varphi^B\!,\delta}(I)\Pi^n_{\rho,\delta},$$
  these operators have the properties
  \begin{align}
    \label{eq:pi:large}
    \tr\,\pi_{I} &\geq 1-2\epsilon, \\
    \label{eq:pi:small}
    \pi_{I}      &\leq \exp\left( -n\bigl(S(\varphi^B|Q)-\delta\bigr) \right)
                             \Pi^n_{\rho,\delta},
  \end{align}
  the latter is obtained by the definition of the conditional typical projector
  in appendix~\ref{app:types}; here, $S(\varphi^B|Q) = \sum_i Q(i)S(\varphi^B_i)$.
  \par
  Denoting the subspace onto which $\Pi^n_{\rho,\delta}$ projects by ${\cal T}$,
  its dimension, by eq.~(\ref{eq:typical:upper}) is bounded
  \begin{equation}
    \label{eq:typical-dim}
    D:=\dim{\cal T} \leq \exp\left( n\bigl(S(\rho)+\delta\bigr) \right).
  \end{equation}
  Now, for the Haar measure ${\rm d}U$ on the unitaries on ${\cal T}$,
  $$\int {\rm d}U\, U\pi_{I}^\top U^* = (\tr\,\pi_{I})\frac{1}{D}\Pi.$$
  Draw $U_1,\ldots,U_K$ i.i.d.~according to the Haar measure. Then, according to
  lemma~\ref{lemma:op:chernoff} stated below,
  \begin{equation*}\begin{split}
    \Pr&\left\{ \frac{1}{K}\sum_k \frac{U_k\pi_{I}^\top U_k^*}{\tr\pi_{I}}
                                       \not\in\frac{1}{D}[(1\pm\epsilon)\Pi] \right\} \\
       &\phantom{===}
        \leq 2D\exp\left( -K\exp\bigl(-n(\chi+2\delta)\bigr)\frac{(1-2\epsilon)\epsilon^2}{2} \right),
  \end{split}\end{equation*}
  with $\chi=S(\rho)-S(\varphi^B|Q)$: because we can rescale the
  $\pi_{I}$ with the factor on the right hand side of eq.~(\ref{eq:pi:small}).
  Thus, by the union bound, there exist $U_1,\ldots,U_K$ such that for
  all $I$ of type $Q$,
  \begin{equation}
    \label{eq:good-choice}
    \frac{1-\epsilon}{D}\Pi \leq \frac{1}{K}\sum_k \frac{1}{\tr\pi_{I}}U_k\pi_{I}^\top U_k^*
                            \leq \frac{1+\epsilon}{D}\Pi,
  \end{equation}
  if $K = (1+n\log m+\log D)\frac{2}{(1-2\epsilon)\epsilon^2}\exp\bigl(n(\chi+2\delta)\bigr)$.
  \par
  The r.s.p.~protocol now works as follows: the Sender, on getting $I$,
  determines its type $Q$ and sends it to the Receiver. If $\|p-Q\|_1 > \delta$,
  the protocol aborts here (this happens with probability $\leq \epsilon$
  if $n$ is sufficiently large, by the law of large numbers).
  For type $Q$ they have agreed on a list of unitaries $U_1,\ldots,U_K$
  as in eq.~(\ref{eq:good-choice}): the Sender can construct the measurement POVM
  \begin{align*}
    A_k             &= \frac{D}{K(1+\epsilon)\tr\pi_{I}} U_k\pi_{I}^\top U_k^*, \\
    A_{\rm failure} &= \Pi - \sum_k A_k,
  \end{align*}
  and measures it (non--destructively) on the maximally
  entangled state $\Phi$ on
  ${\cal T}^A\otimes{\cal T}^B$. The outcome ``${\rm failure}$'' occurs with
  probability less than $\epsilon$, and in the case of outcome $k$
  the Receiver, on learning the value $k$, can apply the
  unitary $U_k^\top$: it is straightforward to check that in this case
  he and the Sender share a purification of $\frac{1}{\tr\pi_{I}}\pi_{I}$.
  Because of eq.~(\ref{eq:pi:large}) and the gentle measurement
  lemma~\ref{lemma:gentle} below, this state has high fidelity to $\varphi^B_{I}$,
  so by~\cite{Uhlmann:fidelity,Jozsa:fidelity}
  she can apply a unitary to her post--measurement state
  to obtain a high--fidelity approximation of $\varphi^{AB}_{I}$.
  \par
  Clearly, this protocol has a high average fidelity. In terms of resources,
  it requires a logarithmic number of bits to communicate the type $Q$ and
  $$\log K\leq n\left( \chi\bigl( \{ Q(i),\varphi_i^B \} \bigr) + f(\delta) \right)$$
  to communicate the result of the measurement described above, with a function
  $f$ which vanishes as $\delta\rightarrow 0$.
  By eq.~(\ref{eq:typical-dim}), it uses
  $$\leq n\left( S\bigl(\sum_i Q(i)\varphi^B_i\bigr) + \delta \right)$$
  ebits. With Fannes inequality~\cite{fannes}
  for $\|p-Q\|_1\leq\delta$, we obtain the claim.
\end{beweis}
\par
\begin{lemma}[``Operator Chernoff bound''~\cite{Ahlswede:Winter}]
  \label{lemma:op:chernoff}
  Let $X_1,\ldots,X_M$ be i.i.d.~random variables taking values in the operators
  ${\cal B}({\cal H})$ on the $D$--dimensional Hilbert space ${\cal H}$,
  $0\leq X_j\leq \1$, with $A=\E X_j\geq\alpha\1$, and let $0<\eta\leq 1/2$. Then
  \begin{equation*}\begin{split}
    \Pr &\left\{ \frac{1}{M}\sum_{j=1}^M X_j \not\in [(1-\eta)A;(1+\eta)A] \right\}        \\
        &\phantom{===============} \leq 2D \exp\left( -M\frac{\alpha\eta^2}{2\ln 2} \right).
  \end{split}\end{equation*}
  \qed
\end{lemma}
\par
\begin{lemma}
  \label{lemma:gentle}
  For a state $\rho$ and an operator $0\leq X\leq\1$, if
  $\tr(\rho X)\geq 1-\epsilon$, then
  $$\left\| \rho-\sqrt{X}\rho\sqrt{X} \right\|_1 \leq \sqrt{8\epsilon}.$$
\end{lemma}
\par
The main result of the present section is that this is essentially optimal:
\begin{thm}
  \label{thm:entg-rsp}
  For the ensemble ${\cal E}=\{ p_i, \varphi_i \}$
  of pure bipartite states and $R\geq 0$,
  $$E^*(R) = N({\cal E},R)
          := \min\{ S(B|C) \,|\, S(X:BC)\leq R \},$$
  where the entropic quantities are with respect to the state
  $\omega$, and minimisation is over all $4$--partite states
  $\omega$ as follows:
  \begin{equation}
    \label{eq:proto}
    \omega = \sum_i p_i\ketbra{i}^X\otimes\varphi_i^{AB}
                              \otimes\sum_j p(j|i)\ketbra{j}^C,
  \end{equation}
  with a classical channel $p(j|i)$.
\end{thm}
\begin{figure}[ht]
  \includegraphics[width=8.5cm]{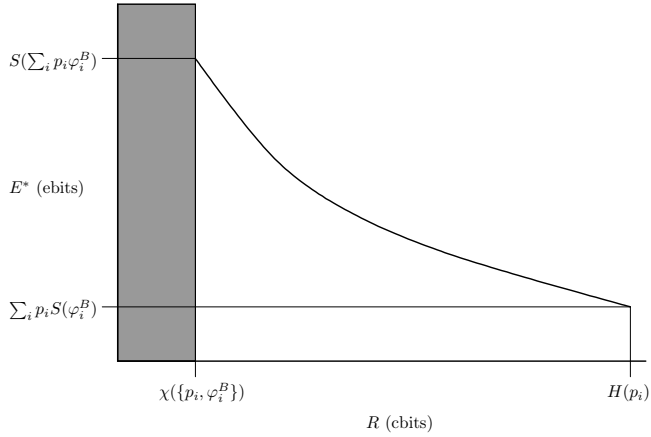}
  \caption{Schematic of the trade--off curve for an ensemble of
    entangled states. The shaded area is forbidden by causality
    and the curve begins at the point
    $\left(\chi\bigl(\{p_i,\varphi_i^B\}\bigr),S\left(\sum_i p_i\varphi_i^B\right) \right)$,
    due to the protocol of proposition~\ref{prop:peters:protocol}. It can never
    go below $E=\sum_i p_i S(\varphi_i^B)$, which is reached at cbit
    rate $R=H(p)$, as this is the very amount of entanglement in the ensemble.}
  \label{fig:ent-schematic}
\end{figure}
\par
This theorem should be compared to the unentangled case,
theorem~\ref{thm:rsp}, to which it provides a pleasingly direct
generalisation. We see that despite the fact that the theorem applies
to ensembles of entangled states, register $A$ does not appear in
any of the entropic quantities involved.
The trade--off curve is a function solely of the ensemble of
mixed states at the Receiver. See Fig.~\ref{fig:ent-schematic}
for a schematic view of the trade--off curve.
\par
Before giving the proof, we state a crucial lemma (compare
lemma~\ref{lemma:N:properties}), which we prove in appendix~\ref{app:proofs}:
\begin{lemma}
  \label{lemma:N-ent:properties}
  $N$ is convex, continuous and strictly decreasing in the interval
  $[S(X:B);S(X)]$. It obeys the following additivity relation
  for ensembles ${\cal E}_1$ and ${\cal E}_2$:
  \begin{equation}
    \label{eq:N-ent:additive}
    N({\cal E}_1\otimes{\cal E}_2,R)
          =\min\bigl\{ N({\cal E}_1 \!,\! R_1)+N({\cal E}_2 \!,\! R_2) | R_1+R_2 \!=\! R \bigr\} \!.
  \end{equation}
  \qed
\end{lemma}
\par
\begin{beweis}[of theorem~\ref{thm:entg-rsp}]
  First, to show that, for fixed $R$, the pair $(R,N(R))$ is achievable, consider
  any channel $q(j|i)$, and let Sender and Receiver perform the following
  procedure (where all information quantities we encounter refer to the
  state $\omega$):
  \par
  In step one, the channel $q^n$ is simulated (using shared randomness)
  on the typical $I$ by the Reverse Shannon Theorem~\cite{BSST,Jozsa:Winter},
  using $n\bigl( I(X:C)+\delta \bigr)$ of forward communication,
  within average total variational distance $\epsilon$ if $n$ is large
  enough.
  \par
  Assuming that the channel $q^n$ is simulated ideally, we can proceed:
  with probability $1-\epsilon$, $J=j_1\ldots j_n$ is typical for the
  distribution $q_j=\sum_i p_i p(j|i)$, i.e. if ${\cal I}_j$ is the set of
  indices $i$ such that $j_i=j$, then
  $$\forall j\quad \bigl| |{\cal I}_j|-nq_j \bigr| \leq \delta.$$
  Now proposition~\ref{prop:peters:protocol} is used to remotely
  prepare the ensemble $\{ q(i|j),\varphi_i \}$ on the block ${\cal I}_j$,
  with the conditional distribution
  $$q(i|j) = \frac{1}{q_j}p_i p(j|i).$$
  This requires
  \begin{align*}
    \leq n(q_j+\delta)\Bigl( \chi(\{q(i|j),\varphi^B_i\})+\delta \Bigr)\text{ cbits,} \\
    \leq n(q_j+\delta)\Bigl( S\bigl(\sum_i q(i|j)\varphi^B_i\bigr)+\delta \Bigr)\text{ ebits.}
  \end{align*}
  In total, we use
  $n\bigl( S(X:C)+S(X:B|C)+f(\delta) \bigr)$ cbits, and
  $n\bigl( S(B|C)+f(\delta) \bigr)$ ebits, and the average fidelity can be
  made arbitrarily close to $1$. Finally, the shared randomness can be disposed of, because
  the average fidelity is an average over it --- hence there exists a value
  of the shared random variable such that the average fidelity is even larger.
  \par
  Now for the converse direction, that $N$ is a lower bound: if $(R,E)$
  is achievable, then for sufficiently large $n$ there exist
  protocols which use $n(R+\delta)$ cbits and $n(E+\delta)$ ebits,
  of fidelity $1-\epsilon$:
  \begin{equation*}
    \sum_I p_I \bra{\varphi_I} \widetilde{\rho}_I \ket{\varphi_I} \geq 1 - \epsilon,
  \end{equation*}
  where $\widetilde{\rho}_I$ is the output state for input $I$.
  Any protocol has the following form: the Sender performs a measurement on
  her half of $n(E+\delta)$ EPR pairs, and then sends $n(R+\delta)$
  bits of classical message $j$ to the Receiver.
  Conditioned on the classical message $j$, he then performs
  a decoding operation $T_j$ on his system.  The outcome is a state
  $\widetilde{\rho}_{I,j}$ such that
  $$\widetilde{\rho}_I = \sum_j p(j|I) \widetilde{\rho}_{I,j}.$$
  \par
  The post--measurement state, including a classical system recording $I$,
  can be written in the form
  \begin{equation*}
    \omega = \sum_I p_I \ketbra{I}^X \otimes \omega_{I,j}^{AB}
                                     \otimes \sum_j p(j|I) \ketbra{j}^C,
  \end{equation*}
  where system $C$ is communicated, and with
  $\widetilde{\rho}_{I,j} = T_j(\omega_{I,j})$.
  By causality, $S_{\omega}(X:BC) \leq n(R+\delta)$.
  Moreover, we can assume that the Receiver's operation $T_j$ does
  not damage the $C$ register since the contents of the register could be
  copied prior to the application of $T_j$.  Since $T_j$ cannot
  increase $S(X:BC)$ by data processing, however, we find that
  for the state
  \begin{equation*}
    \widetilde{\rho} = \sum_I p_I \ketbra{I}^X \otimes \widetilde{\rho}_{I,j}^{AB}
                                               \otimes \sum_j p(j|I) \ketbra{j}^C,
  \end{equation*}
  the inequality
  $$S_{\widetilde{\rho}}(X:BC) \leq n(R+\delta)$$
  holds. Introducing
  $$\Omega = \sum_I p_I \ketbra{I}^X \otimes \varphi_I^{AB}
                                     \otimes \sum_j p(j|I) \ketbra{j}^C,$$
  we conclude that  
  \begin{equation}
    \label{eq:R-bound}
    S_{\Omega}(X:BC) \leq n(R+\delta+f(\epsilon)),
  \end{equation}
  with some universal function $f$ vanishing with $\epsilon$: this is because
  of our fidelity assumption on the protocol and the bilinearity of the
  pure state fidelity, $F(\Omega,\widetilde{\rho}) \geq 1-\epsilon$.
  (Compare the analogous computation in the proof of
  theorem~\ref{thm:rsp}.)
  \par
  To bound the entanglement,
  observe that because the Sender's measurement was on her half of $n(E+\delta)$
  EPR pairs that the state $\omega_j^B = \sum_I q(I|j) \omega_{I,j}^B$
  has support no larger than $2^{n(E+\delta)}$.
  (Note that $p_I p(j|I)$ defines a joint distribution on
  $I$ and $j$.  We use $q(I|j)$ and $q_j$ to denote the associated conditional
  and marginal distributions.)  Therefore, for the state $\omega_{XABC}$,
  \begin{equation*}
    S_{\omega}(B|C) = \sum_j q_j S(\omega_j^B) \leq n(E+\delta).
  \end{equation*}
  If Bob's decoding operation $T_j$ were guaranteed to be unitary we
  could conclude $S_{\widetilde{\rho}}(B|C) \leq n(E+\delta)$.
  More generally, $T_j$ can be decomposed into three steps:
  adjoining an ancilla, applying
  a unitary and then tracing over the ancilla system.  The first two
  steps leave the entropy invariant so without loss of generality, assume
  that conditioned on $j$, Sender and Receiver share a state
  $\omega_{I,j}^{AB B'}$ and that $T_j = \tr_{B'}$.  Our strategy will
  be to use the fact that the states $\varphi_I^{AB}$ are pure to argue
  that the partial trace should not increase the entropy.
  \par
  First, we now have $\omega_{I,j}^{AB} = \widetilde{\rho}_{I,j}^{AB}$.
  Let $\bra{\varphi_I} \widetilde{\rho}_{I,j} \ket{\varphi_I} = 1 - \epsilon_{I,j}$.
  We can choose an extension $\varphi_I^{ABB'}$ of $\varphi_I$ such that
  $F(\varphi_I^{ABB'},\omega_{I,j}^{ABB'})=1-\epsilon_{I,j}$~\cite{Jozsa:fidelity,Uhlmann:fidelity}.
  By the concavity of the fidelity, we then conclude that for
  \begin{align*}
    \varphi_j &:=   \sum_I q(I|j) \varphi_I^{ABB'},   \\
    \omega_j  &:=   \sum_I q(I|j) \omega_{I,j}^{ABB'},
  \end{align*}
  we have
  $$F(\varphi_j,\omega_j) \geq \sum_I q(I|j)(1-\epsilon_{I,j}) =: 1 - \epsilon_j.$$
  Now, because $\varphi_I^{AB}$ is pure, the state $\varphi_j$ must be separable
  across the $AB-B'$ cut.  Therefore, $S(\varphi_j^{BB'}) \geq S(\varphi_j^{B})$.
  On the other hand, using the Fannes inequality and the concavity of its bound,
  we obtain
  \begin{equation*}\begin{split}
    n(E+\delta) &\geq \sum_j q_j S(\omega_j^{BB'})                                \\
                &\geq \sum_j q_j \bigl[S(\varphi_j^{BB'}) - n f(\epsilon_j)\bigr] \\
                &\geq \sum_j q_j S(\varphi_j^{B}) - n f(\epsilon),
  \end{split}\end{equation*}
  for some universal function $f$ vanishing with $\epsilon$.
  Hence,
  \begin{equation}
    \label{eq:E-bound}
    S_\Omega(B|C) \leq n(E+\delta+f(\epsilon)).
  \end{equation}
  Putting this together with eq.~(\ref{eq:R-bound}), we get,
  with $\delta'=\delta+f(\epsilon)$ and the definition of $N$,
  \begin{equation*}
    n(E+\delta') \geq N\bigl( {\cal E}^{\otimes n},n(R+\delta') \bigr).    
  \end{equation*}
  Now we can invoke lemma~\ref{lemma:N-ent:properties}, and obtain
  \begin{equation*}\begin{split}
    E+\delta' &\geq \frac{1}{n} N\bigl( {\cal E}^{\otimes n},n(R+\delta') \bigr) \\
              &=    \min\left\{ \frac{1}{n}\sum_{k=1}^n N({\cal E},R_k)
                                  \left| \frac{1}{n}\sum_{k=1}^n R_k = R+\delta' \right.\right\} \\
              &\geq N({\cal E},R+\delta').
  \end{split}\end{equation*}
  Finally, using continuity of $N$ in $R$, we obtain the result,
  $$E \geq N({\cal E},R).$$
\end{beweis}

\section{Discussion}
\label{sec:conclusion}
In the following subsections we want to review what we have achieved,
while pointing out open questions.

\subsection{Models and resources}
\label{subsec:resources}
In the introduction we have mentioned various subtly different ways
to define remote state preparation (deterministic exact, probabilistic
exact, high fidelity; see next subsection for \emph{oblivious}), as
well as ways to account for the resources used (worst case and expected cost).
\par
Subsequently we have concentrated on probabilistic and high fidelity
asymptotic protocols (for which worst case and expected cost coincide,
as one can easily see). The justification of this choice is that it
seems to be the one best suited to the asymptotic considerations
at our focus.
\par
However, as the following table shows, our conclusions
are for the most part independent of the particulars of the model:
\par\bigskip
\begin{tabular}{l||c|c}
                    & Worst Case                             & Expected           \\
  \hline\hline
  Det. exact        & ?$\geq 1$ ebit, $2\geq$?$\geq 1$ cbits & $1$ ebit, $1$ cbit \\
  \hline
  Prob. exact       &           \multicolumn{2}{c}{$1$ ebit, $1$ cbit}            \\
  \hline
  High fidelity     &           \multicolumn{2}{c}{$1$ ebit, $1$ cbit}            \\
  \hline\hline
  Oblivious         &           \multicolumn{2}{c}{$1$ ebit, $2$ cbits}           \\
  \hline
  Approx. obl.      &           \multicolumn{2}{c}{$1$ ebit, $1$ cbit}
\end{tabular}
\par\bigskip
The entries ``$1$ ebit, $1$ cbit'' derive their achievability from
our protocol $\Pi$ (theorem~\ref{thm:univ:rsp}) --- directly
in the cases ``Probabilistic exact'', ``High fidelity''
and ``Approximately oblivious'' (see the following
subsection), and augmented by teleportation in
the failure event for ``Deterministic exact, Expected cost''.
The upper bound ``$1$ ebit, $2$ cbits'' is of course
teleportation, which indeed is oblivious (see the following
subsection); that in the oblivious case $2$ cbits are indeed
necessary was shown in~\cite{Leung:Shor}.

So, only the entry in the field ``Deterministic exact, Worst case''
is not entirely understood:
in~\cite{HHH:1:2} it is shown that an exact r.s.p.~protocol for a single
qubit requires at least $1$ ebit and $2$ cbits, just like teleportation.
Whether the analogous statement for higher dimensions is true is unknown.

\subsection{Approximate obliviousness}
\label{subsec:oblivious}
An r.s.p. protocol is called oblivious to the Sender~\cite{Leung:Shor}
if, like teleportation, it can be made into a quantum operation for her,
which she can execute without knowing
classically what state she is attempting to prepare.
A protocol is called oblivious to the Receiver~\cite{Leung:Shor}
if, again like teleportation, it leaks no information about the
state being prepared beyond giving him a single specimen of it.
In~\cite{Leung:Shor} it was shown
that if a deterministic exact protocol for preparing states in dimension 
$D$ is oblivious to the Receiver, then it must be oblivious to the Sender
also,  and must therefore, like teleportation,
use at least $\log D$ ebits and $2\log D$ cbits.
\par\medskip
A similar penalty for receiver obliviousness exists even in a purely 
classical analog of r.s.p., namely the simulation of a noisy classical channel by 
noiseless forward classical communication (cbits) and shared randomness (rbits) 
between Sender and Receiver. The classical Reverse Shannon
Theorem~\cite{BSST}
gives a deterministic exact protocol for this task at an expected cbit cost
approaching the simulated channel's classical capacity $C$ in the limit 
of large block size, but it is not hard to show that for some channels any such 
exact efficient simulation must 1) have a worst-case cost exceeding its 
expected cost, and 2) must be non--oblivious to the Receiver.
For example consider a binary
symmetric channel with crossover probability $p$ and capacity
$C=1+p\log p+(1-p)\log(1-p)$. Note that such a channel, given a block of
$n$ inputs, has probability $P_0=(1-p)^n$ of transmitting the whole 
block exactly, without crossovers, and of course any exact
simulation of the channel must simulate
this rare event with the correct probability.  But to avoid a violation 
of causality, the expected cost of the simulation, in instances where
no crossover occurs in a block of size $n$, must be at least
$n-\log(1/P_0)$; otherwise, as in the column method, the Sender could use
$\log(1/P_0)+O(1)$ cbits of additional classical
communication to designate a no--crossover instance within a general 
simulation, thereby communicating $n$ cbits about the input in less
than $n$ cbits of forward communication.
For $0<p<1/2$ the causality--imposed cost
$n-\log(1/P_0)=n(1+\log(1-p)$ exceeds the expected cost $nC$ of an efficient
simulation according to the Reverse Shannon Theorem; therefore in any 
efficient exact simulation, 1) the worst case cost must be at least $n-\log(1/P_0)$;
and 2) the occurrence of a cost exceeding the expected cost $nC$ must be
negatively correlated with the number of crossovers, leaking extra 
information about the channel input besides that contained in the correctly 
simulated output.
\par\medskip
Resuming our discussion of obliviousness in r.s.p., we observe that the previously
studied notions of obliviousness to the Receiver are exact, requiring that the 
protocol leak no information whatever about the input.
In the present paper's main context of approximate simulations it is more appropriate
to use a more robust notion of approximate obliviousness:
\begin{defi}
  \label{defi:approx:oblivious}
  An r.s.p. protocol for a set ${\bf X}$ of states on ${\cal K}$
  is said to be \emph{approximate and approximately oblivious} with
  parameters $(\epsilon,\delta)$ if
  \begin{enumerate}
    \item For all $\sigma\in{\bf X}$, if the Receiver's output state is denoted
          $\rho^B$: $\frac{1}{2}\| \sigma-\rho^B \|_1 \leq \epsilon$.
    \item There exists a c.p.t.p.~map $T$ on the Receiver's system that
          maps his output state $\rho^B$ to a close approximation of
          the whole of what he gets from the protocol: the pre--image of
          $\rho^B$ (under his decoding operation),
          possible residual quantum states, and the classical messages. I.e.,
          $$\frac{1}{2}\| \{\text{Receiver's record}\}-T(\rho^B) \|_1 \leq \delta.$$
  \end{enumerate}
\end{defi}
Note that our notion of ``approximate obliviousness'' does not arise from
some a priori concept of what the Receiver must not learn. It is rather modelled
after ``zero--knowledge'' in zero--knowledge proofs: the verifier gets nothing
that he could not have simulated himself (see~\cite{ZK-Proofs} and subsequent
literature).
\par
Note that for $\epsilon=\delta=0$ we recover the definition of~\cite{Leung:Shor}
of a deterministic exact and exactly oblivious protocol. It would be natural
to conjecture that a robust version of the main result of~\cite{Leung:Shor}
should hold:
\begin{quote}
  For an approximate and approximately oblivious r.s.p.~protocol with parameters
  $(\epsilon,\delta)$ (for the set ${\bf X}={\cal P}({\cal K})$ of all pure states
  on ${\cal K}$), the communication cost is $\geq 2-f(\epsilon,\delta)$ cbits
  per qubit, and it has to use $\geq 1-f(\epsilon,\delta)$ ebits per qubit.
  There, $f$ is a function that vanishes with $\epsilon,\delta\rightarrow 0$.
\end{quote}
Instead, it turns out that our protocol $\Pi$ is indeed
approximate and approximately oblivious in the sense of
definition~\ref{defi:approx:oblivious}, with parameters $(\epsilon,\epsilon)$:
\par
Clearly, part (1) of the definition is satisfied (we remove the failure event
by having the Sender choose one uniformly distributed from the ``good''
messages in the case of a ``failure''). Part (2) also is easily seen to be true:
the simulating map is simply
$$T: \ketbra{\psi} \longmapsto
          \sum_{k=1}^K \frac{1}{K}\ketbra{k} \otimes
                                  \overline{U_k}\ketbra{\psi}U_k^\top.$$
\par
As an aside, we may return to the column--method, presented in
example~\ref{exp:column-method} (without recycling of entanglement):
it is not hard to see that in fact also this
procedure is approximate and approximately oblivious. Indeed, to simulate the
Receiver's view of the protocol, he only has to create an arbitrary state,
say $\left(\frac{1}{D}\1\right)^{\otimes K}$ and an arbitrary classical
message (say, uniformly distributed) with probability $\epsilon$:
this is to simulate the failure. With probability $\frac{1-\epsilon}{K}$ each,
he generates the states
$$\left(\frac{1}{D}\1\right)^{\otimes (k-1)}\otimes\ketbra{\psi}
                                   \otimes\left(\frac{1}{D}\1\right)^{\otimes (K-k)},$$
and the classical message $k=1,\ldots,K$. It is easily seen that
this is $\epsilon$--close to the Receiver's actual view.

\subsection{Further applications of randomisation\protect\\ and trade--off r.s.p.}
\label{subsec:applications}
The remote state preparation of state ensembles turns out to have applications
to other problems, which we simply list here for reference:
\par
1. The protocol we described in section~\ref{sec:ent} for optimal preparation of
pure entangled states produces, when one ignores the Sender's half of the state,
mixed states at the Receiver's system with a classical communication cost
exactly equal to the Holevo quantity of his ensemble.
This result is in fact the
\emph{Quantum Reverse Shannon Theorem}~\cite{QRST} for cq--channels,
and follows also from the alternative protocol
described in~\cite{berry:sanders}.
\par
2. Optimal remote state preparation of entangled states (section~\ref{sec:ent})
is invoked to prove capacity formulas and bounds for the classical communication
capacity of bipartite unitaries assisted by unlimited or bounded
entanglement~\cite{BHLS,harrow}.
\par
3. At the heart of our r.s.p.~protocol is the state randomisation by
relatively few unitaries (theorem~\ref{thm:univ:randomisation}).
In fact, similar to previously considered
\emph{private quantum channels}~\cite{AMTdW,boykin:roychowdhury}
we obtain a private channel scheme, but with halved key length! By applying
the randomisation to half of an entangled state, one even obtains
very efficient schemes for data hiding in bipartite quantum states~\cite{DLT,DHT}.
Our separate paper~\cite{PQC} is devoted to an exploration of these applications.

\acknowledgments
We wish to thank Anura Abeyesinghe, Igor Devetak, Chris Fuchs, Aram Harrow,
Daniel Gottesman and John Smolin for interesting and helpful conversations.
\par
CHB is grateful for the support of the US National Security Agency
and Advanced Research and Development Activity through contracts
DAAD19--01--1--06 and  DAAD19--01--C--0056.
PH acknowledges the support of the Sherman Fairchild Foundation and the US
National Science Foundation under grant no. EIA--0086038.
DL acknowledges the support of the Richard C. Tolman Endowment Fund,
the Croucher Foundation. and the US National Science Foundation
under grant no. EIA--0086038.
AW was supported by the U.K.~Engineering and Physical Sciences Research Council.
PH, DL and AW gratefully acknowledge the hospitality and support of the
Mathematical Sciences Research Institute, Berkeley,
during part of the autumn term of 2002.


\appendix

\section{Gaussian distributed vectors}
\label{sec:gaussian}
This appendix is largely a compilation of known facts about the distribution
of random vectors following a Gaussian law, and of some of their moments:
we freely use textbook knowledge of probability theory (see e.g.~\cite{feller}),
as well as parts of the treatment of large deviation theory by Dembo
and Zeitouni~\cite{Dembo:Zeitouni}.
\par
Recall that the Gaussian (or normal) distribution on the reals with mean $\mu$
and variance $\sigma^2$, denoted $N(\mu,\sigma^2)$, is defined by
the density
$$N(\mu,\sigma^2)\{{\rm d}t\}
       =\frac{1}{\sqrt{2\pi\sigma^2}} e^{-\frac{(t-\mu)^2}{2\sigma^2}}{\rm d}t.$$
We shall phrase most of the following in terms of random variables.
That a random variable $X$ is distributed according to some Gaussian is denoted
$X\sim N(\mu,\sigma^2)$.
\begin{defi}
  \label{defi:complex:gaussian}
  A \emph{Gaussian complex number} with mean $\mu\in\C$ and variance $\sigma^2>0$
  is a random variable $\gamma=X+iY$, where $X$ and $Y$ are independent
  real random variables with
  $X\sim N\!\left({\rm Re}\,\mu,\frac{\sigma^2}{2}\right)$ and
  $Y\sim N\!\left({\rm Im}\,\mu,\frac{\sigma^2}{2}\right)$.
  Its distribution is denoted $N_\C(\mu,\sigma^2)$.
\end{defi}
Note that in this definition we insist that real and imaginary variance
are equal, in contrast to the most general Gaussian distribution in $\R^2$.
\par
Now let ${\cal H}$ be a complex Hilbert space (of finite dimension $d$).
In general, a Gaussian distributed vector is a sum of the form
$\ket{\Gamma}=\sum_j \gamma_j\ket{v_j}$, with an orthonormal basis
$\{\ket{v_j}\}$ and independent Gaussian complex numbers
$\gamma_j\sim N_\C(\mu_j,\sigma_j^2)$. Its distribution is uniquely
determined by the mean $\ket{\mu}=\E\ket{\Gamma}$ and the
\emph{covariance operator} $S^2=\E\ketbra{\Gamma}\geq 0$: the
density is given by
\begin{equation*}
  \Pr \Bigl\{ \ket{\Gamma} \!-\! \ket{\mu}\in \ket{v} \!+\! {\rm d}^{2d}\ket{w} \Bigr\} 
                    \!=\! \frac{1}{\pi^d\det(S^2)}e^{-\bra{v}S^{-2}\ket{v}}{\rm d}^{2d}\ket{w},
\end{equation*}
with the unitarily and translationally invariant normalised volume element
${\rm d}^{2d}w$ in ${\cal H}\simeq\R^{2d}$ (i.e., standard Lebesgue measure).
\par
However, we shall be interested only in the special case that
all means $\mu_j=0$ and all $\sigma_j$ are equal.
\begin{defi}
  \label{defi:vector:gaussian}
  A \emph{symmetric Gaussian vector} with variance $\sigma^2$ is
  a randomly distributed $\ket{\Gamma}\in{\cal H}$ such that
  in one orthonormal basis $\{\ket{v_j}\}$
  $$\ket{\Gamma}=\sum_j \gamma_j\ket{v_j},$$
  with independent $\gamma_j\sim N_\C\!\left(\mu_j,\frac{\sigma^2}{d}\right)$.
  \par
  Equivalently, we could also define it by its covariance operator
  being $\E\ketbra{\Gamma}=\frac{\sigma^2}{d}\1$. From this it follows that
  the distribution of $\Gamma$ is unitarily invariant, hence in the above
  definition we can allow \emph{any orthonormal basis}, a fact we shall make
  frequent use of. Note that $\sigma^2=\E\bra{\Gamma}\Gamma\rangle$.
  This distribution on ${\cal H}$ is denoted $N_{{\cal H}}(0,\sigma^2)$.
\end{defi}
\par
According to Cram\'{e}r's theorem~\cite{Cramer38} (see~\cite{Dembo:Zeitouni}
for its derivation in the present context: it requires only the
``Bernstein trick'' and Markov inequality), for i.i.d.~real random variables
$X,X_1,\ldots,X_N$,
\begin{equation}
  \label{eq:cramer}
  \begin{aligned}
    \Pr\left\{ \frac{1}{N}\sum_{i=1}^N X_i \geq a \right\}
                &\leq \exp\left( -N\,\frac{1}{\ln 2}\inf_{x\geq a}\Lambda^*(x) \right), \\
    \Pr\left\{ \frac{1}{N}\sum_{i=1}^N X_i \leq a \right\}
                &\leq \exp\left( -N\,\frac{1}{\ln 2}\inf_{x\leq a}\Lambda^*(x) \right),
  \end{aligned}
\end{equation}
with the rate function
$$\Lambda^*(x)=\sup_{y\in\R} \left[ yx-\ln\E e^{yX} \right].$$
\par
For a squared Gaussian this can be evaluated explicitly:
\begin{lemma}
  \label{lemma:rate:gaussian}
  For $X=Y^2$, with a Gaussian variable $Y\sim N(0,\sigma^2)$, the rate function
  evaluates to
  \begin{equation*}
    \Lambda^*(x)=\begin{cases}
                   \frac{1}{2}\left[
                                 \frac{x}{\sigma^2}-1-\ln\left(\frac{x}{\sigma^2}\right)
                              \right]    &      :\   x>0,                               \\
                   \infty                &      :\   x\leq 0.
                 \end{cases}
  \end{equation*}
\end{lemma}
\begin{beweis}
  First we calculate $\Lambda(y)=\ln\E e^{yX}$:
  \begin{equation*}\begin{split}
    \E e^{yX} &= \frac{1}{\sqrt{2\pi\sigma^2}}
                    \int_{-\infty}^\infty e^{yt^2}e^{-\frac{t^2}{2\sigma^2}}{\rm d}t \\
            &= \frac{1}{\sqrt{2\pi\sigma^2}}
                  \int_{-\infty}^\infty e^{\left(y-\frac{1}{2\sigma^2}\right)t^2}{\rm d}t \\
            &= \frac{1}{\sqrt{1-2y\sigma^2}}\frac{1}{\sqrt{2\pi\sigma^2}}
                  \int_{-\infty}^\infty e^{-\frac{\tau^2}{2\sigma^2}}{\rm d}\tau          \\
            &= \frac{1}{\sqrt{1-2y\sigma^2}}.
  \end{split}\end{equation*}
  Hence
  \begin{equation*}
    \Lambda(y)=\begin{cases}
                 -\frac{1}{2}\ln(1-2y\sigma^2) & :\ y <  \frac{1}{2\sigma^2}, \\
                 \infty                        & :\ y\geq\frac{1}{2\sigma^2}.
               \end{cases}
  \end{equation*}
  Differentiation reveals one extremum of $yx-\Lambda(y)$
  at $y=\frac{1}{2\sigma^2}-\frac{1}{2x}$,
  which must be the maximum because $yx-\Lambda(y)$ is upper bounded
  for $x>0$ and $-\infty$ at both ends of the permissible interval
  of $y$. This yields the claim.
\end{beweis}
\par
Observe in particular, that $\E X=\sigma^2$, so that we get
for $a=(1+\epsilon)\sigma^2$ and $a=(1-\epsilon)\sigma^2$ ($\epsilon \geq 0$)
in eq.~(\ref{eq:cramer}):
\begin{equation}
  \label{eq:gaussian2:cramer}
  \begin{aligned}
    \Pr\left\{ \frac{1}{N}\sum_{i=1}^N \!X_i \!>\! (1+\epsilon)\sigma^2 \right\}
                    &\!\leq\exp\!\left( -N \frac{\epsilon-\ln(1+\epsilon)}{2\ln 2} \right) \!, \\
    \Pr\left\{ \frac{1}{N}\sum_{i=1}^N \!X_i \!<\! (1-\epsilon)\sigma^2 \right\}
                    &\!\leq\exp\!\left( -N \frac{-\epsilon-\ln(1-\epsilon)}{2\ln 2} \right) \!.
  \end{aligned}
\end{equation}
We shall make use of the following lower bound:
\begin{equation}
  \label{eq:gaussian:rate:lower}
  \text{For all }-1\leq\xi\leq 1,\quad
              \frac{1}{2\ln 2}\bigl(\xi-\ln(1+\xi)\bigr) \geq \frac{\xi^2}{12\ln 2}.
\end{equation}
Proof is by Taylor expansion: for $|\xi|=1$ it is obviously true, and for
$|\xi|<1$ we have
\begin{equation*}\begin{split}
  \xi-\ln(1+\xi) &= \xi-\left( \sum_{n=1}^\infty (-1)^{n-1}\frac{\xi^n}{n} \right)   \\
                &=  \sum_{n=2}^\infty (-1)^n\frac{\xi^n}{n}                          \\
               &=   \sum_{k=1}^\infty \left[\frac{\xi^{2k}}{2k}-\frac{\xi^{2k+1}}{2k+1}\right] \\
              &\geq \sum_{k=1}^\infty \left[\frac{\xi^{2k}}{2k}-\frac{\xi^{2k}}{2k+1}\right]
                                                                           \geq \frac{\xi^2}{6}.
\end{split}\end{equation*}

\section{State randomisation}
\label{sec:randomisation}
\begin{beweis}[of lemma~\ref{lemma:conc}]
  Since the Haar measure is left and right invariant, we may assume that
  $\varphi = \ketbra{e_1}$ and $P = \sum_{i=1}^p \ketbra{e_i}$ for some fixed
  orthonormal basis $\{\ket{e_i}\}$.
  Let $\ket{\Gamma_j} = \sum_{i=1}^D g_{ij} \ket{e_i}$,
  where $g_{ij} \sim N_\C(0,1)$ are i.i.d. (see appendix~\ref{sec:gaussian}).
  The distribution of $\ket{\Gamma_j}$
  is the same as the distribution for $\| \Gamma_j \|_2 U \ket{e_1}$
  if $U$ is chosen using the Haar measure.
  \par
  For fixed $U = U_j$ and $\ket{\Gamma}=\ket{\Gamma_j}$, the convexity of
  $\exp$ implies that
  \begin{equation*}\begin{split}
    \E_\Gamma \exp &\left( \frac{y}{D} \sum_{i=1}^p | \bra{e_i} \Gamma\rangle |^2 \right) \\
                   &=    \E_U \E_\Gamma \exp\left( \frac{y \| g \|_2^2}{D}
                                           \sum_{i=1}^p | \bra{e_i}U\ket{e_1} |^2 \right) \\
                   &\geq \E_U \exp\left( \E_\Gamma \frac{y \| g \|_2^2}{D}
                                           \sum_{i=1}^p | \bra{e_i}U\ket{e_1} |^2 \right) \\
                   &=    \E_U \exp\left( y \sum_{i=1}^p |\bra{e_i}U\ket{e_1}|^2 \right)   \\
                   &=    \E_U \exp\bigl( y \tr(U \varphi U^* P) \bigr).
  \end{split}\end{equation*}
  Invoking Cram\'{e}r's theorem, this inequality between the
  moment generating functions establishes that
  $\frac{1}{n}\sum_{j=1}^n \tr(U\varphi U^* P)$
  converges to its mean value
  \begin{equation*}
    \E_U \tr(U\varphi U^* P) =
    \E_\Gamma \frac{1}{D} \sum_{i=1}^p |\bra{e_i}\Gamma\rangle|^2 = \frac{p}{D}
  \end{equation*}
  at least as quickly as
  $$\frac{1}{n}\sum_{j=1}^n \sum_{i=1}^p \frac{1}{D} | \bra{e_i}\Gamma_j\rangle |^2
                     = \frac{1}{n} \sum_{j=1}^n \sum_{i=1}^p \frac{1}{D} |g_{ij}|^2.$$
  That is, the exponential rate function $\Lambda^*_U$ controlling large deviations
  of $\tr(U\varphi U^* P)$ is
  at least as large as the corresponding function $\Lambda^*_\Gamma$
  for $\frac{1}{D} \sum_{i=1}^p |\bra{e_i}\Gamma\rangle|^2$.
  \par
  The latter we have evaluated and estimated in section~\ref{sec:gaussian}:
  if $|\epsilon|\leq 1$, $\Lambda^*_\Gamma(1+\e) \geq \frac{1}{6}\epsilon^2 p$
  and the result follows by an application of the union bound.
\end{beweis}
\par
\begin{beweis}[of lemma~\ref{lemma:net}]
  We begin by relating the trace norm to the Hilbert space norm:
  \begin{equation*}\begin{split}
    \bigl\| \ket{\psi}-\ket{\varphi} \bigr\|_2^2
                      &=    2-2{\rm Re}\,\bra{\psi}\varphi\rangle  \\
                      &\geq 2-2|\bra{\psi}\varphi\rangle|          \\
                      &=    2\left(1-\sqrt{F(\psi,\varphi)}\right) \\
                      &\geq 1-F(\psi,\varphi)                      \\
                      &\geq \left( \frac{1}{2}\bigl\| \psi-\varphi \bigr\|_1 \right)^2,
  \end{split}\end{equation*}
  where the last line is a well--known relation between fidelity
  and trace norm distance~\cite{Fuchs:vandeGraaf}.
  Thus it will be sufficient to find an $\epsilon/2$--net for the Hilbert
  space norm. Let ${\cal M} = \{\ket{\varphi_i}:1\leq i \leq m\}$ be a maximal set
  of pure states satisfying $\| \ket{\varphi_i} - \ket{\varphi_j} \|_2 \geq \epsilon/2$
  for all $i$ and $j$. By definition, ${\cal M}$ is an $\epsilon/2$--net for 
  $\|\cdot\|_2$. We can estimate $m$ by a volume argument, however. As
  subsets of $\R^{2D}$, the open balls of radius $\epsilon/4$ about each
  $\ket{\varphi_i}$ are pairwise disjoint and all contained in the ball
  of radius $1+\epsilon/4$ centered at the origin. Therefore,
  \begin{equation*}
    m (\epsilon/4)^{2D} \leq \left(1+\epsilon/4\right)^{2D},
  \end{equation*}
  and we are done.
\end{beweis}

\section{Universal quantum--classical state description}
\label{sec:universal:qc-tradeoff}
In section~\ref{sec:optimality} we reduced universal r.s.p. with little
entanglement resources to universal visible quantum data compression with
the same amount of qubit resources. Here we study the latter question.
\par
For a Hilbert space ${\cal K}$ of dimension $D$
a (universal) \emph{quantum--classical state compression}
(or \emph{quantum--classical state description}) of fidelity $F$
consists of the following: first, a map
$$E:\psi \longmapsto E(\psi)=(\xi(\psi),m(\psi)),$$
mapping every pure state vector $\ket{\psi}\in{\cal K}$ to a pair $(\xi,m)$,
where $\ket{\xi}\in{\cal C}$ is a state vector in the \emph{(quantum) code space}
and $m$ is a classical message from the set ${\cal M}$.
Second, a family of completely positive and trace preserving linear maps
$$D_m:{\cal B}({\cal C}) \longrightarrow {\cal B}({\cal K}),$$
such that
$$\forall\ket{\psi}\in{\cal K}\quad
    F\left( \psi,D_{m(\psi)}\bigl(\xi(\psi)\bigr) \right) \geq F.$$
We call such a compression/description ``universal'' because it has to have
high fidelity for every possible input pure state. Note that both the quantum and
classical parts of the state description are of fixed size, in contrast to variable--length
coding schemes existing in classical and quantum data compression,
for which the qualifier ``universal'' has a quite different meaning: there it means
that the encoding of a state has the minimal possible length according to some
standard. Here we are interested in how the two resources we have trade against
each other, in a ``universal'' way.
\par
There are two extreme examples. One is ``no classical message'', i.e. $|{\cal M}|=1$
and a $D$--dimensional ${\cal C}\simeq{\cal K}$:
for this the Sender simply prepares the desired state $\psi$ in ${\cal C}$.
On the other end, $\dim{\cal C}=1$ (i.e., no quantum  message), in which case
one can achieve fidelity $1-\epsilon$ by identifying an element of
an $\epsilon$--net ${\cal M}$ in ${\cal K}$: by
lemma~\ref{lemma:net} this requires $\left(4+\log\frac{1}{\epsilon}\right)D$ cbits.
\par
The following theorem says that there occurs a jump in going from one extreme
to the other, in the sense that as soon as the quantum resources are less than $\log D$
qubits, an exponential number of classical bits are needed:
\begin{thm}
  \label{thm:univ:description}
  A quantum--classical state compression with average fidelity $F$:
  $$\int {\rm d}\ket{\psi}\, F\left( \psi,D_{m(\psi)}\bigl(\xi(\psi)\bigr) \right) \geq F,$$
  which uses a code space ${\cal C}$ of dimension $S\leq qD$ ($q<F$),
  requires exponential classical resources:
  $$\log|{\cal M}| \geq \frac{q(1-q)}{6}D-2\log D+\log\left(1-\sqrt{\frac{1-F}{1-q}}\right).$$
\end{thm}
\begin{beweis}
  Write the fidelity $F=1-\epsilon$ and define
  $${\cal S}_m:=\left\{ \ket{\psi}\in{\cal K} | \exists\ket{\phi}\in{\cal C}\
                                     F\bigl(T_m(\phi),\psi\bigr) \geq 1-\vartheta \right\},$$
  the set of pure states which can be reached to fidelity $1-\vartheta$ using
  the message $m$ and some quantum code state. Clearly,
  $\Sigma := \bigcup_{m\in{\cal M}} {\cal S}_m$ is the set of all states which
  can be decoded with fidelity $1-\vartheta$. By Markov's inequality,
  $$\lambda(\Sigma) \geq 1-\frac{\epsilon}{\vartheta} = 1-\frac{1}{\sqrt{t}},$$
  where $\lambda$ is the unique ${\cal U}(D)$--invariant measure on pure states,
  normalised to $1$ (i.e., a probability measure), and
  with $t:=\frac{1-q}{1-F}>1$ and $\vartheta=\sqrt{t}\epsilon$.
  \par
  Hence, to prove a lower bound on
  $|{\cal M}|$, it will be sufficient to prove an upper bound on the volume
  $\lambda({\cal S}_m)$ of the sets ${\cal S}_m$.
  \par
  We concentrate on a particular message $m$ for the time being, so we drop the
  subscript $m$ in the sequel.
  The decoding operation $T:{\cal B}({\cal C})\rightarrow{\cal B}({\cal K})$
  can be written, by a result of Choi~\cite{Choi}, as
  $$T(\phi)=\sum_{i=1}^{D^2} A_i \phi A_i^*,$$
  with linear operators $A_i:{\cal C}\rightarrow{\cal K}$.
  Hence we can write
  $$T(\phi)=\sum_{i=1}^{D^2} p_i\phi_i,$$
  with probabilities $p_i$ and pure state vectors $\ket{\phi_i}\in A_i{\cal C}=:{\cal W}_i$,
  the latter an (at most) $S$--dimensional subspace of ${\cal K}$.
  But if $F(T(\phi),\psi)\geq 1-\epsilon$, there must exist $i$ such that
  $F(\phi_i,\psi)\geq 1-\epsilon$, by bilinearity of the pure state fidelity.
  \par
  Hence,
  \begin{equation}
    \label{eq:covering}
    {\cal S}\subset\bigcup_{i=1}^{D^2} B_\epsilon({\cal W}_i),
  \end{equation}
  with
  $$B_\epsilon({\cal W}):=\left\{ \ket{\psi} | \exists\ket{\phi}\in{\cal W}\
                                     |\langle\psi\ket{\phi}|^2\geq 1-\epsilon \right\},$$
  and it will be sufficient to bound the volume of $B_\epsilon({\cal W})$ for an
  arbitrary $S$--dimensional subspace ${\cal W}$:
  \par
  Denoting the orthogonal projector onto ${\cal W}$ by $P$, we can rewrite
  $B_\epsilon({\cal W})$ as
  $$B_\epsilon({\cal W})=\left\{ \ket{\psi} | \tr(\ketbra{\psi}P) \geq 1-\epsilon \right\}.$$
  Also, since the volume $\lambda$ is a probability measure, we have
  \begin{equation*}\begin{split}
    \lambda(B_\epsilon({\cal W}))
               &=\Pr\left\{ \ket{\psi} | \tr(\ketbra{\psi}P) \geq 1-\epsilon \right\}  \\
               &=\Pr\left\{ U | \tr\bigl(U\ketbra{0}U^*P\bigr) \geq 1-\epsilon \right\},
  \end{split}\end{equation*}
  with ${\cal U}(D)$--uniformly distributed unit vector $\ket{\psi}$
  and a unitary $U$ distributed according to Haar measure.
  Observing that the expectation of the overlap $\tr(\ketbra{\psi}P)$
  above is $q$, and defining
  $$\eta := \min\left\{ \frac{1-\vartheta}{q}-1,1 \right\} \geq \sqrt{t\epsilon},$$
  we can use lemma~\ref{lemma:conc} to bound this probability by
  $$\exp\left( -qD\frac{\eta^2}{6} \right),$$
  so using the union bound in eq.~(\ref{eq:covering}) we have
  $$\lambda({\cal S}) \leq D^2\exp\left( -qD\frac{\eta^2}{6} \right),$$
  which implies what we wanted:
  $$\log|{\cal M}| \geq \frac{q\eta^2}{6}D-2\log D+\log\left(1-\sqrt{\frac{1-F}{1-q}}\right).$$
\end{beweis}
\begin{rem}
  \label{rem:sharp:result}
  There exists a universal quantum--classical state compression with
  fidelity $\geq (1-\epsilon)^2$, which uses a code space ${\cal C}$ of dimension
  $S=\left\lceil \left(1-\frac{\epsilon}{2}\right)D \right\rceil$ and classical
  communication of $\left\lceil \epsilon^{-1} \right\rceil$ cbits.
  \par
  This works as follows: decompose ${\cal K}$ into orthogonal subspaces ${\cal H}_k$
  ($k=0,\ldots,K=\left\lceil \epsilon^{-1} \right\rceil$), such that
  $$\dim{\cal H}_0 < \dim{\cal H}_1 = \ldots = \dim{\cal H}_K
                     = \left\lfloor \frac{D}{K} \right\rfloor.$$
  Write $P_k$ for the projectors onto the orthogonal complement of ${\cal H}_k$:
  then
  $$\frac{1}{K}\sum_{k=1}^K P_k \geq \left( 1-\frac{1}{K} \right)\1
                                \geq (1-\epsilon)\1,$$
  which means that for every state vector $\psi$ the Sender can find
  $1\leq k\leq K$ such that $\tr\bigl(\ketbra{\psi}P_k\bigr) \geq 1-\epsilon$.
  The Sender simply transmits the projected quantum state and $k$,
  from which the Receiver can reconstruct $\psi$ to the desired fidelity.
  The rank of the $P_k$ determines $S$, which is easily estimated.
  \qed
\end{rem}
This result is in contrast to the findings of~\cite{HJW}, where for the
asymptotic compression of longer and longer products of qubits (or qu--$d$--its
in general) a rate trade--off between qubits and cbits was exhibited.
In the light of the present theorem we can understand how that comes about:
the model of~\cite{HJW} admits only \emph{product states} in larger and larger
spaces. The trade--off curve then quantifies how efficiently the manifold of
product states can be covered by (neighbourhoods of) small subspaces.
\par
Once we admit all states in dimension $D$, this covering, instead of using
polynomially (in $D$) many subspaces, requires exponentially (in $D$) many!

\section{Typical subspaces}
\label{app:types}
The following material can be found in most textbooks on information
theory, e.g.~\cite{cover:thomas,csiszar:koerner}, or in the original
literature on quantum information
theory~\cite{Schumacher,Schumacher:Jozsa,SW:coding,winter:qstrong}.
\par
For strings of length $n$ from a finite alphabet ${\cal X}$, which we generically
denote $x^n=x_1\ldots x_n\in{\cal X}^n$, we define the \emph{type of $x^n$} as
the empirical distribution of letters in $x^n$: i.e., $P$ is the type of $x^n$ if
$$\forall x\in{\cal X}\quad P(x)=\frac{1}{n}|\{ k:x_k=x \}|.$$
It is easy to see that the total number of types is upper bounded by
$(n+1)^{|{\cal X}|}$.
\par
The \emph{type class of $P$}, denoted ${\cal T}_P^n$, is defined as all
strings of length $n$ of type $P$. Obviously, the type class is obtained by
taking all permutations of an arbitrary string of type $P$.
\par
The following is an elementary property of the type class:
\begin{equation}
  \label{eq:type:cardinality}
  (n+1)^{-|{\cal X}|}\exp\bigl( nH(P) \bigr)
                        \leq |{\cal T}_P^n| \leq \exp\bigl( nH(P) \bigr),
\end{equation}
with the (Shannon) entropy $H(P)$.
\par
For $\delta>0$, and for an arbitrary probability
distribution $P$, define the set of \emph{$P$--typical sequences} as
$${\cal T}^n_{P,\delta} := \left\{ x^n: \left|-\frac{1}{n}\log P^{\otimes n}(x^n)-H(P)\right|
                                                                                \leq \delta\right\}.$$
By the law of large numbers, for every $\epsilon>0$ and sufficiently large $n$,
\begin{equation}
  \label{eq:typical:prob}
  P^{\otimes n}({\cal T}^n_{P,\delta}) \geq 1-\epsilon.
\end{equation}
Furthermore:
\begin{align}
  \label{eq:typical:upper}
  |{\cal T}^n_{P,\delta}| &\leq \exp\bigl( n(H(P)+\delta) \bigr), \\
  \label{eq:typical:lower}
  |{\cal T}^n_{P,\delta}| &\geq (1-\epsilon)\exp\bigl( n(H(P)-\delta) \bigr).
\end{align}
\par
For a (classical) channel $W:{\cal X}\longrightarrow{\cal Y}$ (i.e. a stochastic map, taking
$x\in{\cal X}$ to a probability distribution $W_x$ on ${\cal Y}$) and a string
$x^n\in{\cal X}^n$ of type $P$ we denote the output distribution of $x^n$ in
$n$ independent uses of the channel by
$$W^n_{x^n} = W_{x_1}\otimes\cdots\otimes W_{x_n}.$$
Let $\delta>0$, and define the set of \emph{conditional $W$--typical sequences} as
$${\cal T}^n_{W,\delta}(x^n) := \left\{ y^n: \left|-\frac{1}{n}\log W^n_{x^n}(y^n)-H(W|P)\right|
                                                                                \leq \delta\right\},$$
where $H(W|P)=\sum_x P(x)H(W_x)$ is the conditional entropy.
\par
Once more by the law of large numbers, for every $\epsilon$ and
sufficiently large $n$,
\begin{equation}
  \label{eq:c-typical:prob}
  W^n_{x^n}\bigl( {\cal T}^n_{W,\delta}(x^n) \bigr) \geq 1-\epsilon.
\end{equation}
Furthermore:
\begin{align}
  \label{eq:c-typical:upper}
  \bigl| {\cal T}^n_{W,\delta}(x^n) \bigr| &\leq \exp\bigl( n(H(W|P)+\delta) \bigr), \\
  \label{eq:c-typical:lower}
  \bigl |{\cal T}^n_{W,\delta}(x^n) \bigr| &\geq (1-\epsilon)\exp\bigl( n(H(W|P)-\delta) \bigr).
\end{align}
\par
All of these concepts and formulas have analogues as ``typical projectors'' $\Pi$ for
quantum state: by virtue of the spectral decomposition, the eigenvalues of a density
operator can be interpreted as a probability distribution over eigenstates. The subspaces
spanned by the typical eigenstates are the ``typical subspaces''. The trace of a density
operator with one of its typical projectors is then the probability of the corresponding
set of typical sequences.
\par
Notations like $\Pi^n_{\rho,\delta}$, $\Pi^n_{\varphi,\delta}(i^n)$ etc. for
a state $\rho$ and a cq--channel $\varphi$ should be clear from this.
\par
There is only one such statement for density operators that we shall use, which is
not of this form:
\begin{lemma}[Operator law of large numbers]
  \label{lemma:law:large-numbers}
  Let $x^n\in{\cal X}^n$ be of type $P$, and let $W:{\cal X}\longrightarrow{\cal S}({\cal H})$
  be a cq--channel. Denote the average output state of $W$ under $P$ as
  $$\rho = \sum_x P(x) W_x.$$
  Then, for every $\epsilon>0$ and sufficiently large $n$,
  $$\tr\bigl( W^n_{x^n}\Pi^n_{\rho,\delta} \bigr)  \geq 1-\epsilon.$$
\end{lemma}
\begin{beweis}
  See~\cite{winter:qstrong}, Lemma 6.
\end{beweis}

\section{A possible operational reduction of r.s.p.~to q.c.t.}
\label{app:operational}
Our protocol in section~\ref{sec:tradeoff} for (asymptotic) remote state
preparation of ensembles reduces the problem to the quantum--classical
trade--off in visible source coding~\cite{HJW} by an operational reduction:
we simply add our universal r.s.p.~protocol, theorem~\ref{thm:univ:rsp},
on top of the q.c.t.~coding, theorem~\ref{thm:qct}.
The optimality proof,
though modelled closely along the lines of the corresponding proof
in~\cite{HJW}, is however completely independent.
It would be desirable to have a closer connection between the trading
of qubits vs.~cbits and of ebits vs.~cbits, and in this appendix we describe
an operational link going the other way, from r.s.p.~to q.c.t., resting on an
(as yet unproven) conjecture on mixed--state compression:
\par
More precisely, given an r.s.p.~protocol
(asymptotic and approximate) of cbit rate $C$ and ebit rate $E$
construct a q.c.t.~scheme with cbit rate $R=C-E$ and qubit
rate $Q=E$. This would exactly revert the construction of
section~\ref{sec:tradeoff}.
\par
We will prove that this is possible, assuming
the following conjecture (see~\cite{BCFJS} and~\cite{Jozsa:Winter}):
\begin{conj}
  \label{conj:mixed:CR}
  Given an i.i.d.~source ${\cal F}=\{p_i,\rho_i\}$ of \emph{mixed states}
  it is possible to visibly compress the source asymptotically and approximately,
  using shared randomness between Sender and Receiver, and
  communicating qubits at rate
  $$\chi\bigl( \{p_i;\rho_i\} \bigr) = S\left(\sum_i p_i\rho_i\right)-\sum_i p_i S(\rho_i).$$
\end{conj}
Note that this is true if the ensemble consists of pure states,
by Schumacher's quantum data compression~\cite{Schumacher}.
Also observe that the conjecture certainly is true for commuting mixed
states: this is essentially the content of the Reverse Shannon
Theorem~\cite{BSST}, see also~\cite{Jozsa:Winter}.
\par
Note (as we have observed earlier) that shared randomness can safely
be assumed free, because we are considering an average pure state fidelity
as quality measure of the protocol.
\par
We assume the following general form of our r.s.p.~protocol:
it uses a standard maximally entangled state $\ket{\Phi}$ on
${\cal K}_A\otimes{\cal K}_B$,
with $\dim{\cal K}\leq 2^{n(E+\delta)}$. Depending on $I=i_1\ldots i_n$
the Sender makes a measurement on ${\cal K}_A$, described by a
POVM ${\bf A}^{(I)}=\left(A^{(I)}_m\right)$, where $m$
is the message she subsequently sends to the Receiver, chosen from a set of
$M\leq 2^{n(R+\delta)}$. Of course, as $n$ tends to infinity, $\delta$
will tend to zero. For each of the messages $m$, the Receiver can execute
an operation $T_m$ on ${\cal K}_B$, acting on the state induced
by the entanglement and the measurement, together with the outcome,
denoted $\rho_{m|I}$. Denote the induced probability of the
message $m$ (given $I$) as $q(m|I)$. We shall only assume
the ``local'' fidelity condition, eq.~(\ref{eq:local-fidelity}),
not the stronger ``global'' one, eq.~(\ref{eq:global-fidelity}).
\par
Our goal is to re--enact the creation of the post--measurement state
and the transmission of the classical message using only cbit and
qubit communications. The key idea comes from the observation that
there is noise in the system due to the uncontrollable randomness
of the POVMs. We want to transfer the generation of this noise to
the shared randomness.
\par
We shall now look at blocks formed from the $n$--blocks given by the
assumed r.s.p.~protocol. We use the previous notation
$I=i_1\ldots i_n$ for an $n$--block, and introduce $I^N=I_1\ldots I_N$
for such a block of blocks.
By the Reverse Shannon Theorem (in the formulation of~\cite{Jozsa:Winter})
we can visibly encode the distribution $q(\cdot|I^N)$,
at least for typical $I^N$, using shared randomness and communicating
\begin{equation*}\begin{split}
  {\rm I}(I : m) &=    H(m)-\sum_I p_I H\bigl(q(\cdot|I)\bigr) \\
                 &\leq n(R+\delta)-\sum_I p_I H\bigl(q(\cdot|I)\bigr)
\end{split}\end{equation*}
cbits per $n$--block, where we treat $I$ and $m$ as jointly distributed
random variables:
$$\Pr\{ I,m \} = p_I q(m|I),$$
with $H$ is the usual Shannon entropy, and ${\rm I}$ the Shannon
mutual information.
\par
A feature of the Reverse Shannon Theorem that was noted earlier is that
the Sender gets full feedback, i.e.~she obtains the very (random)
message $m$ the Receiver gets out.
With the help of this feedback, she just prepares the post--measurement state
on ${\cal K}$ that otherwise the Receiver would have found on his half of
the entanglement, and sends it. Then, obviously, the Receiver can proceed
as in the r.s.p.~protocol. It is clear, that we end up with a procedure
having high fidelity according to the ``local'' fidelity criterion
eq.~(\ref{eq:local-fidelity}), now over a block of length $Nn$.
\par
How does this behave in terms of resources? Clearly, we now use only qubits
and cbits. Inspection of the above formulas reveals that all is fine if
\begin{equation}
  \label{eq:crux}
  \sum_I p_I H\bigl(q(\cdot|I)\bigr) \geq n(E-\delta'),
\end{equation}
with $\delta'=o(1)$ as $n\rightarrow\infty$. Because then we
have a q.c.t.~scheme (satisfying eq.~(\ref{eq:local-fidelity}))
that uses $Nn(E+\delta)$ qubits and $Nn(R-E+\delta+\delta')$
cbits. This is exactly the reduction we wanted: since
in~\cite{HJW} the trade--off curve was (implicitly)
proved for the criterion eq.~(\ref{eq:local-fidelity}), we obtain
the desired bounds on $E$ and $R$.
\par
We are left with proving that assuming the negation of
eq.~(\ref{eq:crux}) leads to a contradiction:
so, introducing the tripartite state
$$\omega = \sum_I p_I \ketbra{I}^A \otimes \sum_m q(m|I)\rho_{m|I}^B\otimes\ketbra{m}$$
for notational convenience,
assume that there exists $\Delta E>0$ such for all large $n$
\begin{equation}
  \label{eq:absurd}
  S(B:C|A) \leq S(C|A) \leq n(E-\Delta E).
\end{equation}
The right inequality is the negation of eq.~(\ref{eq:crux}). and
the left is by data processing: for each value of $I$ in $A$
the information between $B$ and $C$ (which is the Holevo quantity
of the ensemble $\{q(\cdot|I),\rho_{\cdot|I}\}$) is upper bounded
by the entropy of $C$, i.e.~$H(q(\cdot|I))$.
\par
Note further that, because $\sum_m q(m|I)\rho_{m|I}$ equals the
maximally mixed state for all $I$, we have $S(A:B)=0$, hence by
the chain rule for quantum mutual information,
$$S(B:C|A)=S(AC:B).$$
\par
Thus, for large enough $N$, we can, by conjecture~\ref{conj:mixed:CR},
encode $N$--blocks of the $\rho_{m|I}$ using shared randomness and sending
\begin{equation*}\begin{split}
  N S(AC:B)+o(N) &=    N S(B:C|A)+o(N) \\
                 &\leq Nn(E-\Delta E)+o(N)
\end{split}\end{equation*}
qubits: the conjecture is applied to the ensemble $\{p_I q(m|I);\rho_{m|I}\}$,
which partly is given (the input, $I$) and which partly is obtained by simulating
the noisy classical channel $q(\cdot|\cdot)$ (the variable $m$).
Observe that $m$ is by this method
generated simultaneously at the Sender and at the Receiver.
\par
Switching back to r.s.p.~via eq.~(\ref{eq:E:upperbound})
we end up with a protocol on $Nn$--blocks using
only $Nn(E-\Delta E)+o(N)$ ebits and
\begin{equation*}\begin{split}
  N S(A:C) &+ N S(B:C|A) + o(N) \\
           &= N S(AB:C) + o(N) \leq Nn(R+o(1))
\end{split}\end{equation*}
cbits: the first term is due to the communication cost of the Reverse Shannon
Theorem, and the second is the cost overhead to remotely prepare the
$N S(B:C|A)+o(N)$ qubits of the compressed mixed states.
In the limit this leads to a rate pair $(R,E-\Delta E)$,
contradicting the optimality of $(R,E)$.
\qed

\section{Miscellaneous proofs}
\label{app:proofs}
\begin{beweis}[of lemma~\ref{lemma:N:properties}]
  For finiteness of the values of $N$ we have to have $R\geq S(A:B) = S(B)$,
  which is clearly sufficient. For $N({\cal E},R)=0$ on the other hand, one
  has to have a state with $0=S(A:B|C)=S(B|C)$. But then,
  \begin{equation*}\begin{split}
    R &\geq S(A:BC) \\
      &=    S(A:C) + S(A:B|C) = S(A:C).
  \end{split}\end{equation*}
  However, $S(B|C)=0$ says that $B$ is in a pure state given $C$,
  which is only possible if $S(A|C)=0$. Hence $R\geq S(A)$, which
  clearly is sufficient, too.
  \par
  For convexity, let $\omega_1$ be optimal for $R_1$, $\omega_2$
  optimal for $R_2$, i.e.
  \begin{equation*}\begin{split}
    S_{\omega_k}(A:B|C) &= N(R_k), \\
    S_{\omega_k}(A:BC)  &\leq R_k,
  \end{split}\end{equation*}
  $k=1,2$. Furthermore, let $0\leq \lambda\leq 1$. Then form the state
  $$\omega = \lambda\omega_1\otimes\ketbra{1}^{C'}
            +(1-\lambda)\omega_2\otimes\ketbra{2}^{C'}.$$
  By definition (with $\widetilde{C}=CC'$),
  \begin{align*}
    S_\omega(A:B\widetilde{C})  &\leq \lambda R_1 + (1-\lambda)R_2, \\
    S_\omega(A:B|\widetilde{C}) &=    \lambda N(R_1) + (1-\lambda) N(R_2),
  \end{align*}
  and thus the minimisation yields
  $$N\bigl( \lambda R_1 + (1-\lambda)R_2 \bigr) \leq \lambda N(R_1) + (1-\lambda) N(R_2).$$
  \par
  Taking $R_1=S(B)$ and $R_2=S(A)$, we obtain that in the interval
  $[S(B);S(A)]$ is strictly decreasing and continuous --- otherwise there
  were a contradiction to convexity. (Note that $N(R_2)=0$!)
  \par
  Finally, for the additivity relation,
  eq.~(\ref{eq:N:add}), observe that ``$\leq$'' is almost obvious:
  if $\omega_k$ are optimal for $({\cal E}_k,R_k)$, $k=1,2$, it is immediate to
  check that $\omega=\omega_1\otimes\omega_2$ is feasible for
  $({\cal E}_1\otimes{\cal E}_2,R=R_1+R_2)$, implying an upper bound
  of $N({\cal E}_1,R_1)+N({\cal E}_2,R_2)$ for $N({\cal E}_1\otimes{\cal E}_2,R)$.
  \par
  In the other direction, let $\omega$ be optimal for $({\cal E}_1\otimes{\cal E}_2,R)$:
  $$\omega \!=\! \sum_{i,i'}\! p_i p_{i'}' \ketbra{i}^{A_1}\otimes\ketbra{i'}^{A_2}
                                           \pi_i^{B_1}\otimes\pi_{i'}^{B_2}
                                           \otimes\sum_j\! p(j|ii')\ketbra{j}^C \!.$$
  First, by the chain rule and data processing,
  \begin{equation*}\begin{split}
    R &\geq S(A_1A_2:B_2B_2C)                   \\
      &=    S(A_1:B_1B_2C) + S(A_2:B_1B_2C|A_1) \\
      &\geq S(A_1:B_1C)    + S(A_2:B_2C|A_1).
  \end{split}\end{equation*}
  Thus we can write $R=R_1+R_2$ such that
  \begin{equation}
    \label{eq:R1R2}
    S(A_1:B_1C)\leq R_1,\quad S(A_2:B_2C|A_1)\leq R_2.
  \end{equation}
  Second, by a similar reasoning,
  \begin{equation*}\begin{split}
    N({\cal E}_1\otimes{\cal E}_2,R) &=    S(A_1A_2:B_2B_2|C)                   \\
                                     &=    S(A_1:B_1B_2|C) + S(A_2:B_1B_2|CA_1) \\
                                     &\geq S(A_1:B_1|C)    + S(A_2:B_2|CA_1).
  \end{split}\end{equation*}
  Here, the first term is $\geq N({\cal E}_1,R_1)$ by definition, using
  eq.~(\ref{eq:R1R2}). The second term is similarly $\geq N({\cal E}_2,R_2)$,
  using additionally the convexity of $N$.
\end{beweis}
\par\medskip
\begin{beweis}[of lemma~\ref{lemma:N-ent:properties}]
  Monotonicity follows directly from the definition.
  \par
  For finite values we obviously have to have
  $$R\geq S(X:BC) \geq S(X:B).$$
  Also always (using that the conditional entropy can only increase under
  quantum operations --- a consequence of strong subadditivity),
  $$E\geq S(B|C) \geq S(B|X),$$
  with equality when $C$ contains a copy of $X$.
  \par
  Convexity is proved exactly as in the proof of lemma~\ref{lemma:N:properties}.
  From this continuity in the domain of finite values follows, as well as
  strict monotonicity as long as $N(R) > S(B|X)$.
  \par
  It remains to prove the additivity relation, eq.~(\ref{eq:N-ent:additive}):
  ``$\leq$'' is the trivial inequality, after the pattern of
  the proof of lemma~\ref{lemma:N:properties}.
  As for ``$\geq$'', consider an optimal state $\omega$ for
  ${\cal E}_1\otimes{\cal E}_2$ and rate $R$:
  \begin{equation*}\begin{split}
    \omega = \sum_{i,i'} p_i p_{i'}\ketbra{i}^{X_1}\otimes\ketbra{i'}^{X_2}
                             &\otimes\varphi_i^{A_1B_1}\otimes\varphi_{i'}^{A_2B_2} \\
                             &\otimes\sum_j p(j|ii')\ketbra{j}^C.
  \end{split}\end{equation*}
  Then, using the chain rule, data processing, and the independence
  of $X_1$ and $X_2$,
  \begin{equation*}\begin{split}
    R &\geq S(X_1X_2:B_1B_2C)                   \\
      &=    S(X_1:B_1B_2C) + S(X_2:B_1B_2C|X_1) \\
      &\geq S(X_1:B_1C)    + S(X_2:B_2C|X_1)    \\
      &=    S(X_1:B_1C)    + S(X_2:B_2CX_1),
  \end{split}\end{equation*}
  so we can find $R_1$ and $R_2$ such that $R_1+R_2=R$ and
  $$S(X_1:B_1C)\leq R_1,\quad S(X_2:B_2CX_1)\leq R_2.$$
  On the other hand,
  \begin{equation*}\begin{split}
    N({\cal E}_1\otimes{\cal E}_2) &=    S(B_1B_2|C)              \\
                                   &=    S(B_1|C) + S(B_2|CB_1)   \\
                                   &\geq S(B_1|C) + S(B_2|CX_1),  \\
                                   &\geq N({\cal E}_1,R_1) + N({\cal E}_2,R_2)
  \end{split}\end{equation*}
  where in the third line we have used that conditional entropy can only increase
  under quantum operations, a consequence of strong subadditivity. The last line
  follows because with our choice of $R_1$ and $R_2$, $C$ and $CX_1$ are permitted
  in the definition of $N({\cal E}_1,R_1)$ and $N({\cal E}_2,R_2)$, respectively.
\end{beweis}


\begin{thebibliography}{M}
  \bibitem{Ahlswede:Winter} R. Ahlswede, A. Winter, ``Strong Converse for Identification
    Via Quantum Channels'', IEEE Trans. Inf. Theory, vol. 48, no. 3, pp. 569--579, 2002.
    Addendum \emph{ibid.}, vol. 49, no. 1, p. 346, 2003.

  \bibitem{AMTdW} A. Ambainis, M. Mosca, A. Tapp, R. de Wolf, ``Private Quantum Channels'',
    Proc. $41^{\rm st}$ FOCS, pp. 547--553, IEEE Computer Society Press, 2000.

  \bibitem{BCFJS} H. Barnum, C. M. Caves, C. A. Fuchs, R. Jozsa, B. W. Schumacher,
    ``On quantum coding for ensembles of mixed states'', J. Phys. A: Math. and Gen.,
    vol. 34, no. 35, pp. 6767--6785, 2001.

%
  \bibitem{Dense:coding} C. H. Bennett, S. Wiesner, ``Communication via one-- and
    two--particle operators on Einstein--Podolsky--Rosen states'', Phys. Rev. Letters,
    vol. 69, 2881--2884, 1992.

  \bibitem{Teleportation} C. H. Bennett, G. Brassard, C. Cr\'{e}peau, R. Jozsa,
    A. Peres, W. K. Wootters, ``Teleporting an unknown quantum state via dual classical
    and Einstein--Podolsky--Rosen channels'', Phys. Rev. Letters, vol. 70, no. 13,
    pp. 1895--1899, 1993.

  \bibitem{QRST} C. H. Bennett, I. Devetak, A. Harrow, P. W. Shor, A. Winter,
    ``The Quantum Reverse Shannon Theorem'', in preparation.

  \bibitem{BDSSTW} C. H. Bennett, D. P. DiVincenzo, P. W. Shor, J. A. Smolin,
    B. M. Terhal, W. K. Wootters, ``Remote State Preparation'', Phys. Rev. Letters,
    vol. 87, 077902, 2001. (Erratum \emph{ibid.}, vol. 88, 099902, 2002.)

  \bibitem{BHLS} C. H. Bennett, A. W. Harrow, D. W. Leung, J. A. Smolin,
    ``On the capacities of bipartite Hamiltonians and unitary gates'',
    e--print {\tt quant-ph/0205057}, 2002.

  \bibitem{PQC} C. H. Bennett, P. Hayden, D. W. Leung, P. W. Shor, A. Winter,
    ``Randomizing quantum states: Constructions and applications'',
    e--print {\tt quant-ph/0307xxx}, 2003.

  \bibitem{BSST} C. H. Bennett, P. W. Shor, J. A. Smolin, A. V. Thapliyal,
    ``Entanglement--Assisted Classical Capacity of Noisy Quantum Channels'', Phys.
    Rev. Letters, vol. 83, no. 15, pp. 3081--3084, 1999, and
    ``Entanglement--assisted capacity of a quantum channel and the reverse
    Shannon theorem'', IEEE Trans. Inf. Theory, vol. 48, no. 10,
    pp. 2637--2655, 2002.

  \bibitem{berry:sanders} D. W. Berry, B. C. Sanders, ``Optimal Remote State
    Preparation'', Phys. Rev. Letters, vol. 90, 057901, 2003.

  \bibitem{boykin:roychowdhury} P. O. Boykin, V. Roychowdhury, ``Optimal encryption
    of quantum bits'', Phys. Rev. A, vol. 67, 042317, 2003.

  \bibitem{CGM} N. Cerf, N. Gisin, S. Massar, ``Classical Teleportation
    of a Quantum Bit'', Phys. Rev. Letters, vol. 84, no. 11, pp. 2521--2524, 2000.

  \bibitem{Chernoff} H. Chernoff, ``A measure of asymptotic efficiency for tests of a hypothesis
    based on the sum of observations'', Ann. Math. Statistics, vol. 23, pp. 493--507, 1952.

  \bibitem{Choi} M.-D. Choi, ``Completely positive linear maps on complex matrices'',
    Linear Algebra and Appl., vol. 10, pp. 285--290, 1975.

  \bibitem{cover:thomas} T. Cover, J. A. Thomas, \emph{Elements of Information Theory},
    John Wiley \& Sons, Inc., New York, 1991.

  \bibitem{Cramer38} H. Cram\'{e}r, ``Sur un nouveau th\'{e}or\`{e}me--limite de
    la theorie des probabilit\'{e}s'', Actualit\'{e}s Scientifiques et Industrielles,
    no. 736 (Colloque consacr\'{e} \`{a} la theorie des probabilit\'{e}s),
    pp. 5--23, Hermann, Paris, 1938.

  \bibitem{csiszar:koerner} I. Csisz\'{a}r, J. K\H{o}rner, \emph{Information Theory: Coding
    Theorems for Discrete Memoryless Systems},
    Academic Press, Inc., New York--London, 1981.

%
  \bibitem{davies:lewis} E. B. Davies, J. T. Lewis, ``An operational approach to
    quantum probability'', Comm. Math. Phys., vol. 17, pp. 239--260, 1970.

  \bibitem{Dembo:Zeitouni} A. Dembo, O. Zeitouni,
    \emph{Large Deviations: Techiques and Applications}, 2nd edition,
    Springer Verlag (Series Applications of Mathematics 38), New York, 1998.

  \bibitem{Devetak:Berger} I. Devetak, T. Berger, ``Low--Entanglement Remote State
    Preparation'', Phys. Rev. Letters, vol. 87, 197901, 2001.

  \bibitem{DLT} D. P. DiVincenzo, D. W. Leung, B. M. Terhal, ``Quantum Data Hiding'',
     IEEE Trans. Inf. Theory, vol. 48, no. 3, pp. 580--599, 2002.

  \bibitem{DHT} D. P. DiVincenzo, P. Hayden, B. M. Terhal, ``Hiding Quantum Data'',
    e--print {\tt quant-ph/0207147}, 2002.

  \bibitem{fannes} M. Fannes, ``A continuity property of the entropy density for
    spin lattice systems'', Comm. Math. Phys., vol. 31, pp. 291--294, 1973.

  \bibitem{feller} W. Feller, \emph{An introduction to probability theory and its
    applications}, Vol.~I, $3^{\rm rd}$ ed., John Wiley \& Sons,
    New York--London--Sydney, 1968.

  \bibitem{Fuchs:vandeGraaf} C. A. Fuchs, J. van de Graaf, ``Cryptographic
    distinguishability measures for quantum-mechanical states'', IEEE Trans.
    Inf. Theory, vol. 45, no. 4, pp. 1216--1227, 1999.

  \bibitem{ZK-Proofs} S. Goldwasser, S. Micali, C. Rackoff, ``The Knowledge Complexity
    of Interactive Proof Systems'', SIAM J. Comput., vol. 18, no. 1, pp. 186--208, 1989.

  \bibitem{HHH:1:2} A. Hayashi, T. Hashimoto, M. Horibe, ``Remote State Preparation
    Without Oblivious Condition'', Phys. Rev. A, vol. 67, no. 5, 052302, 2003.

  \bibitem{HJW} P. Hayden, R. Jozsa, A. Winter, ``Trading quantum for classical
    resources in quantum data compression'', J. Math. Phys., vol. 43,
    no. 9, pp. 4404--4444, 2002.

  \bibitem{harrow} A. Harrow, ``Coherent Classical Communication'',
    e--print {\tt quant-ph/0307091}, 2003.

  \bibitem{Jozsa:fidelity} R. Jozsa, ``Fidelity for mixed quantum states'',
    J. Mod. Optics, vol. 41, no. 12, pp. 2315--2323, 1994.

  \bibitem{Kushilevitz:Nisan} E. Kushilevitz, N. Nisan, \emph{Communication Complexity},
    Cambridge University Press, 1996.

  \bibitem{Schumacher:Jozsa} R. Jozsa, B. Schumacher, ``A new proof of the quantum
    noiseless coding theorem'', J. Mod. Optics, vol. 41, no. 12, pp. 2343--2349, 1994.

  \bibitem{Leung:Shor} D. W. Leung, P. W. Shor, ``Oblivious remote state preparation'',
    Phys. Rev. Letters, vol. 90, no. 12, 127905, 2003.

  \bibitem{Lo1999} H.--K. Lo, ``Classical Communication Cost in Distributed Quantum
    Information Processing --- A generalization of Quantum Communication Complexity'',
    Phys. Rev. A, vol. 62, 012313, 2000.

  \bibitem{Pati} A. K. Pati, ``Minimum classical bit for remote preparation and measurement
    of a qubit'', Phys. Rev. A, vol. 63, 014302, 2001.

  \bibitem{Schumacher} B. W. Schumacher, ``Quantum Coding'', Phys. Rev. A, vol. 51,
    no. 4, pp. 2738--2747, 1995.

  \bibitem{SW:coding} B. Schumacher, M. D. Westmoreland, ``Sending classical information
    via noisy quantum channels'', Phys. Rev. A, vol. 56, no. 1, pp. 131--138, 1997.

  \bibitem{Uhlmann:fidelity} A. Uhlmann, ``The `transition probability' in the state
    space of a ${}^*$--algebra'', Rep. Math. Phys., vol. 9, no. 2,
    pp. 273--279, 1976.

  \bibitem{winter:qstrong} A. Winter, ``Coding theorem and strong converse for
    quantum channels'', IEEE Trans. Inf. Theory, vol. 45, no. 7, pp. 2481--2485 , 1999.

  \bibitem{Jozsa:Winter} A. Winter, ``Compression of sources of probability
    distributions and density operators'', e--print {\tt quant-ph/0208131}, 2002.

  \bibitem{Zeng:Zhang} B. Zeng, P. Zhang, ``Remote--state preparation in higher dimension
     and the parallelizable manifold $S^{n-1}$'', Phys. Rev. A, vol. 65, 022316, 2002.

\end{thebibliography}
\end{document}